\documentclass[fleqn,usenatbib]{mnras}

\usepackage{graphicx}	% Including figure files
\usepackage{amsmath}	% Advanced maths commands
\usepackage{txfonts}
\usepackage[T1]{fontenc}

\PassOptionsToPackage{unicode}{hyperref}
\PassOptionsToPackage{naturalnames}{hyperref}
\usepackage{hyperref}

\DeclareRobustCommand{\VAN}[3]{#2}
\let\VANthebibliography\thebibliography
\def\thebibliography{\DeclareRobustCommand{\VAN}[3]{##3}\VANthebibliography}

\usepackage{todonotes}
\newcommand{\Mth}{M_\mathrm{th}}
\newcommand{\Mp}{M_\mathrm{p}}
\newcommand{\Hp}{H_\mathrm{p}}
\newcommand{\hp}{h_\mathrm{p}}
\newcommand{\Rp}{R_\mathrm{p}}
\newcommand{\lsh}{l_\mathrm{sh}}
\newcommand{\OmK}{\Omega_\mathrm{K}}
\newcommand{\Omp}{\Omega_\mathrm{p}}
\newcommand{\Fdep}{F_\mathrm{dep}}
\newcommand{\fdep}{f_\mathrm{dep}}
\newcommand{\Pp}{P_\mathrm{p}}
\newcommand{\md}{\mathrm{d}}
\newcommand{\Sigmap}{\Sigma_\mathrm{p}}
\newcommand{\tgap}{t_\mathrm{gap}}
\newcommand{\Sigmad}{\Sigma_\mathrm{d}}

\newcommand{\appropto}{\mathrel{\vcenter{
  \offinterlineskip\halign{\hfil$##$\cr
    \propto\cr\noalign{\kern2pt}\sim\cr\noalign{\kern-2pt}}}}}

\title[Early stages of gap opening]{
Early stages of gap opening by planets in protoplanetary discs
}

\author[Cordwell \& Rafikov]{
Amelia J. Cordwell,$^{1}$\thanks{E-mail: ajc356@cam.ac.uk}
Roman R. Rafikov$^{1, 2}$
\\
$^{1}$Department of Applied Mathematics and Theoretical Physics, University of Cambridge, Wilberforce Road, Cambridge CB3 0WA, UK\\
$^{2}$Institute for Advanced Study, Einstein Drive, Princeton, NJ 08540, USA
}

\date{Accepted XXX. Received YYY; in original form ZZZ}

\pubyear{2023}

\begin{document}
\label{firstpage}
\pagerange{\pageref{firstpage}--\pageref{lastpage}}
\maketitle

\begin{abstract}
%%%  This is 250 words
Annular substructures in protoplanetary discs, ubiquitous in sub-mm observations, can be caused by gravitational coupling between a disc and its embedded planets. Planetary density waves inject angular momentum into the disc leading to gap opening only after travelling some distance and steepening into shocks (in the absence of linear damping); no angular momentum is deposited in the planetary coorbital region, where the wave has not shocked yet. Despite that, simulations show mass evacuation from the coorbital region even in inviscid discs, leading to smooth, double-trough gap profiles. Here we consider the early, time-dependent stages of planetary gap opening in inviscid discs. We find that an often-overlooked contribution to the angular momentum balance caused by the time-variability of the specific angular momentum of the disc fluid (caused, in turn, by the time-variability of the radial pressure support) plays a key role in gap opening. Focusing on the regime of shallow gaps with depths of $\lesssim 20\%$, we demonstrate analytically that early gap opening is a self-similar process, with the amplitude of the planet-driven perturbation growing linearly in time and the radial gap profile that can be computed semi-analytically. We show that mass indeed gets evacuated from the coorbital region even in inviscid discs. This evolution pattern holds even in viscous discs over a limited period of time. These results are found to be in excellent agreement with 2D numerical simulations. Our simple gap evolution solutions can be used in studies of dust dynamics near planets and for interpreting protoplanetary disc observations.
\end{abstract}


%%%%%%%%%%%%%%%%%%%%%%%%%%%%%%%%%%%%%%%%%%%%%%%%%%

%%%%%%%%%%%%%%%%% BODY OF PAPER %%%%%%%%%%%%%%%%%%

%%%%%%%%%%%%%%%%%%%%%%%%%%%%%%%%%%%%%%%%%%%%%%%%%%
%%%%%%%%%%%%%%%%%%%%%%%%%%%%%%%%%%%%%%%%%%%%%%%%%%

\section{Introduction}
\label{sec:intro}

%%%%%%%%%%%%%%%%%%%%%%%%%%%%%%%%%%%%%%%%%%%%%%%%%%

Over the last decade, resolved observations of protoplanetary discs have shown that annular substructures such as rings and gaps are ubiquitous in these discs \citep{andrews_observations_2020}. While their origin is still debated, a promising possibility is the gravitational effect of massive perturbers --- planets orbiting within the disc \citep{Lin1986,Dong2017,bae_formation_2017,zhang_disk_2018}. The gravity of such an embedded planet excites density waves \citep{goldreich_excitation_1979,GT1980} which travel in a spiral pattern \citep{ogilvie_wake_2002,rafikov_nonlinear_2002} across the disc until they dissipate. After the wave dissipates its angular momentum gets transferred to the disc fluid, leading to the formation of gaps in surface density \citep{rafikov_planet_2002}. 

While some wave dissipation can occur  due to viscosity \citep{Takeuchi1996}, this is of secondary importance to dissipation due to the non-linear evolution and shocking of the wave \citep{goodman_planetary_2001, rafikov_nonlinear_2002} or linear wave damping due to radiative cooling \citep{miranda_planet-disk_2020,miranda_gaps_2020}. In inviscid discs with no cooling, the only possible cause of dissipation is the non-linear evolution of the planet-driven density wave. This is the situation that we will focus on in this work. Planets with masses $\Mp$ below the thermal mass \citep{goodman_planetary_2001},
\begin{equation}
    \Mth = \left(\frac{\Hp}{\Rp}\right)^3 M_{\star},
\end{equation}
where $\Hp$ is the local scale height of the disc at the radius $\Rp$ of the planetary orbit and $M_{\star}$ is the mass of the star, excite waves that do not begin to shock (and therefore transfer their angular momentum to the disc) until they travel a shocking length  \citep{goodman_planetary_2001},
\begin{equation}
    \lsh \approx 0.8 \Hp \left(\frac{\gamma + 1}{12/5} \frac{\Mp}{\Mth}\right)^{-2/5},
    \label{eq:lsh}
\end{equation}
radially away from the planet \citep{goodman_planetary_2001}, where $\gamma$ is the adiabatic index. 

\begin{figure}
\centering
\includegraphics[width=0.48\textwidth]{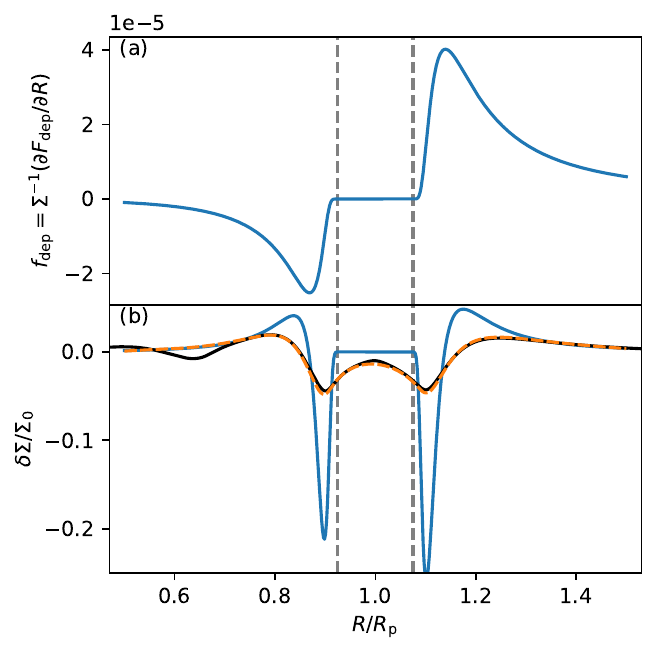}
\caption{
(a) Representative model of the angular momentum deposition torque density $\partial\Fdep/\partial R$ (normalised by $\Sigma$, in arbitrary units), equation (\ref{eq:amfd_func}), evaluated for a $\Mp/\Mth = 0.25$ planet in a globally isothermal disc with local scale height $h_p = 0.03$ and background surface density slope of $p=1.5$. Note that $\partial\Fdep/\partial R=0$ in the coorbital region. (b) Relative surface density deviation $\delta\Sigma/\Sigma_0$ obtained at $t = 100 \Pp$ using different methods: blue --- via equation (\ref{eq:no-ldot-soln}) not accounting for the $\dot l$ term, orange --- via equation (\ref{eq:global_inviscid}) incorporating the $\dot l$ term, black --- taken from a full numerical simulation (agreeing well with the orange curve near the planet). Gray vertical dashed lines are at one shocking length away from the planet, $|R-\Rp|=\lsh$, marking the coorbital region.  }
\label{fig:fig_one}
\end{figure}

Very importantly, in this picture there is no deposition of angular momentum in the co-orbital region of the disc near the planet, at $|R-\Rp|<\lsh$. This is illustrated in Figure \ref{fig:fig_one}a where we show the deposition torque density $\partial \Fdep/\partial R$ --- the amount of angular momentum deposited by the planet-driven wave per unit radial distance. One can see  that indeed $\partial \Fdep/\partial R=0$ for $|R-\Rp|<\lsh$. One might then naively expect that the disc surface density would not evolve in that region \citep{rafikov_planet_2002}. This expectation is illustrated by the blue curve in Figure \ref{fig:fig_one}b, which shows the surface density deviation $\delta\Sigma/\Sigma$ and is computed after 100 orbits based on the standard assumption \citep{Lynden1974} that the mass flux in the disc is determined solely by the angular momentum deposition (shown in panel (a)) into the disc fluid, see Section \ref{sec:initial_evolution} for details. Indeed, this curve shows no deviation of $\Sigma$ from its initial value in the co-orbital region, while also exhibiting two deep gaps just beyond $\lsh$ (where $|\partial \Fdep/\partial R|$ peaks on each side of the planet, as predicted in \citet{rafikov_planet_2002}.

Despite this logic, hydrodynamic simulations of inviscid discs, with no radiative or viscous wave damping, do show a significant surface density evolution in the coorbital region  \citep{duffell_global_2012,zhu_low-mass_2013,Zhu2014,Dong2017}. This is illustrated in Figure \ref{fig:fig_one}b by the black curve based on our inviscid simulations (see Section \ref{sec:results}), which differs from the blue curve in several important ways. First and foremost, $\Sigma$ deviates from its initial value in the co-orbital region, $\delta\Sigma/\Sigma<0$ for $|R-\Rp|<\lsh$, contrary to the earlier expectation. Second, outside the coorbital region ($|R-\Rp|>\lsh$) the amplitude of $\delta\Sigma/\Sigma$ variations is significantly lower than that exhibited by the blue curve, although the 'double gap' structure still remains in place. Third, the black curve experiences a small dip around $0.65\Rp$, which is absent in the blue curve. 

We therefore have a qualitative discrepancy that must be investigated \citep{muto_two-dimensional_2010}: why is mass evacuated from the co-orbital region even though there is no angular momentum injection in that part of the (inviscid) disc, and why has the earlier theory not captured this phenomenon? This issue is less of a problem in viscous discs, as viscosity smooths out radial inhomogeneities of the surface density and can wipe out the belt of gas at $\Rp$ squeezed between the two growing gaps on each side of the planet. In the long run, a balance between the dissipation torque density and the viscous stress will be established leading to a steady state gap around the planetary orbit \citep{duffell_empirically_2020, kanagawa_formation_2015, fung_how_2014}. However, as viscosity is now believed to be small in many discs \citep{rafikov_protoplanetary_2017,Flaherty2017,Flaherty2018}, the inviscid approximation may be more relevant, and in many systems the gaps we observe may not be in a steady state.

%%%%%%%%%%%%%%%%%%%%%%%%%%%%%%%%%%%%%%%%%%%%%%%%%%
%%%%%%%%%%%%%%%%%%%%%%%%%%%%%%%%%%%%%%%%%%%%%%%%%%

In this work we address the aforementioned issues by focusing on the simplest case of a low-mass, non-migrating planet in an invisicd disc to re-evaluate the connection between angular momentum deposition and surface density evolution. Additional effects such as planetary accretion and migration, as well as the long-term disc evolution can also impact the development of a gap by a planet \citep{rafikov_planet_2002,Nazari2019}, however, to untangle their contribution it is essential we have a solid understanding of gap formation in the simplest setup. We will demonstrate that the mass evacuation from the co-orbital region is explained by accounting for a certain term in the angular momentum balance equation, proportional to the time derivative of the specific angular momentum of the disc fluid, that is often neglected in studies of astrophysical discs. We show that in the linear regime, for small gap depths, one can derive an analytical solution for the evolution of gap profile in time and space (equations (\ref{eq:full_solution}), (\ref{eq:global_solution})), given the knowledge of an angular momentum deposition function $\partial \Fdep/\partial R$. Throughout this work we will compare our models to hydrodynamical simulations.

Our work is organized as follows. In Section \ref{sec:physical_setup} we describe our physical setup, and in Section \ref{sec:basic_equations} we re-derive the evolution of the disc surface density due to angular momentum injection inncluding the aforementioned time-dependent term. In Section \ref{sec:initial_evolution} we apply this theory in the linear regime to derive an analytical solution for the growth of a gap in the local (Section \ref{sec:local}) and global (Appendix \ref{sec:app-non-loc}) settings. In Sections \ref{sec:results}, we compare our analytical solution to inviscid numerical simulations, while in Section \ref{sec:viscosity} we extend our analysis to viscous discs and compare our findings with viscous simulations. Finally, in Section \ref{sec:discussion} we describe the applications (including observational), limitations and possible extensions of this work and summarize our findings in Section \ref{sec:sum}.

%%%%%%%%%%%%%%%%%%%%%%%%%%%%%%%%%%%%%%%%%%%%%%%%%%
%%%%%%%%%%%%%%%%%%%%%%%%%%%%%%%%%%%%%%%%%%%%%%%%%%

\section{Physical setup}
\label{sec:physical_setup}

%%%%%%%%%%%%%%%%%%%%%%%%%%%%%%%%%%%%%%%%%%%%%%%%%%

We consider an initially axisymmetric, two-dimensional (2D) protoplanetary disc with a radially varying background (unperturbed) surface density $\Sigma_0(R)$ in the form of a power law: 
\begin{align}
    \Sigma_0(R) = \Sigma_\mathrm{p} \left(\frac{R}{\Rp} \right)^{-p}.
    \label{eq:Sig-PL}
\end{align}
In our derivations we generally allow for a locally isothermal equation of state (EoS) $P = c_s^2(R) \Sigma$ where $c_s(R)$ is a radially-varying sound speed, but most of our results are obtained for the globally isothermal EoS for which $c_s$ is a constant. The globally isothermal EoS is chosen since the locally isothermal EoS is known to not conserve angular momentum flux of the density waves \citep{miranda_gaps_2020}, which would complicate the analysis of the problem. The scale height of the disc is defined as $H = c_s/\OmK$ where $\OmK(R)$ is the Keplerian angular velocity due to the central star and $h \equiv H/R$ is the disc aspect ratio. 

A planet with mass below the thermal mass, $\Mp < \Mth$, is placed in a fixed circular orbit at $R = \Rp$ inside the disc, with Keplerian angular velocity $\Omp = \sqrt{GM_{\star}/\Rp^3}$. Also, $\Pp=2\pi/\Omp$ is the oribtal period of the planet and $\hp$ is the disc aspect ratio at $\Rp$. In our analytical calculations (Section \ref{sec:basic_equations}) the deposition of angular momentum into the disc fluid through weakly non-linear shocks driven by a planet is modeled using the semi-analytical approach based on \cite{cimerman_planet-driven_2021}. 

We consider both inviscid (Sections \ref{sec:results} and \ref{sec:initial_evolution}) and viscous discs (Section \ref{sec:viscosity}). For viscous discs the kinematic viscosity is given by
\citep{shakura_black_1973}
\begin{equation}
    \nu = \alpha c_s H = \alpha \frac{c_s^2}{\OmK},
    \label{eq:nu}
\end{equation}
where $\alpha$ is a constant. In viscous discs we set the background surface density of the disc such that prior to the introduction of the planet the surface density is in a viscous steady state with mass accretion rate $\dot M\propto\nu \Sigma = \text{const}$, i.e. $\Sigma_0 \propto R^{-1.5}$ for globally isothermal discs.

Our results are presented in units of $G = M_{\star} = \Rp = 1$ and normalised against the background surface density $\Sigma_0$.

%%%%%%%%%%%%%%%%%%%%%%%%%%%%%%%%%%%%%%%%%%%%%%%%%%
%%%%%%%%%%%%%%%%%%%%%%%%%%%%%%%%%%%%%%%%%%%%%%%%%%

\section{Basic equations}
\label{sec:basic_equations}

%%%%%%%%%%%%%%%%%%%%%%%%%%%%%%%%%%%%%%%%%%%%%%%%%%

To describe the planet-driven evolution of a disc we start from the standard one-dimensional (1D) equations of continuity and angular momentum conservation \citep{pringle_accretion_1981, rafikov_planet_2002, muto_two-dimensional_2010}: 
\begin{align}
    \label{eq:base_mass}
    R \frac{\partial \Sigma}{\partial t} + \frac{\partial}{\partial R}(R \Sigma v_R) &= 0,\\
    \label{eq:base_amf}
    R\frac{\partial}{\partial t}(\Sigma R^2 \Omega) + \frac{\partial}{\partial R}(R \Sigma v_R \Omega R^2) &= -\frac{1}{2 \pi}\left( \frac{\partial G}{\partial R} - \frac{\partial \Fdep}{\partial R} \right),
\end{align}
where $\Omega$ is the gas angular velocity, and $v_R$ is its radial velocity, both of which are properly averaged, see below. Also,
$G \equiv - 2 \pi R^3 \nu \Sigma \left(d \Omega/d R\right)$ is the viscous angular momentum flux. For completeness, in Appendix \ref{sec:3d} we provide a derivation of these 1D equations from the full 2D continuity and Navier-Stokes equations, showing that they reduce to the form (\ref{eq:base_mass}), (\ref{eq:base_amf}) exactly provided that $\Sigma$ and the velocity components have been azimuthally averaged and mass-weighted, i.e.
\begin{align}    
\langle \Sigma \rangle  \equiv \frac{1}{2 \pi} \oint \Sigma d \phi, 
%    \label{eq:avSig}\\
~~~~~~~~~~    \langle v_i \rangle  \equiv \frac{1}{2 \pi \langle \Sigma \rangle } \oint v_i \Sigma d \phi. 
    \label{eq:avv}
\end{align}
In particular, angular velocity is defined as $\Omega=\langle v_\phi\rangle/R$. In equations (\ref{eq:base_mass}), (\ref{eq:base_amf}) we drop $\langle..\rangle$ for brevity. 

The effect of planet driven waves is included through the $\partial \Fdep/\partial R$ term, which represents the deposition of the wave angular momentum into the disc. Introducing the wave angular momentum flux (AMF)
\begin{equation}
    \label{eq:f_wave}
    F_{wave}(R) \equiv R^2 \oint \Sigma(R, \phi) v_R(R, \phi) \delta v_{\phi}(R, \phi) d\phi,
\end{equation}
and the gravitational torque density on the disc due to the planetary potential $\Phi_\mathrm{p}({\bf R})$,
\begin{equation}
    \frac{\partial T}{\partial R} \equiv R \oint \Sigma(R, \phi) \frac{\partial \Phi_\mathrm{p}}{\partial \phi} d\phi,
    \label{eq:torque}
\end{equation}
conservation of angular momentum allows us to express
\begin{equation}
    \frac{\partial \Fdep}{\partial R} =\frac{\partial T}{\partial R} -  \frac{\partial F_{wave}}{\partial R}.
    \label{eq:Fdep}
\end{equation}

Combining equations (\ref{eq:base_mass}) and (\ref{eq:base_amf}) one obtains the radial velocity
\begin{equation}
    v_R = -\frac{1}{R \Sigma} \left(\frac{\partial l}{\partial R} \right)^{-1} \left[\frac{1}{2 \pi}\left(\frac{\partial G}{\partial R} - \frac{\partial \Fdep}{\partial R} \right) + \Sigma R \frac{\partial l}{\partial t} \right],
    \label{eq:vR}
\end{equation}
where $l = R^2 \Omega$ is the specific angular momentum of the disc fluid. Substituting this into equation (\ref{eq:base_mass}) we find the 1D equation for the evolution of surface density:
\begin{equation}
    \label{eq:time_evolve_sigma}
    \frac{\partial \Sigma}{\partial t}  = \frac{1}{R} \frac{\partial }{\partial R} \left\{ \left( \frac{\partial l}{\partial R}\right)^{-1} \left[ \frac{1}{2 \pi} \left(\frac{\partial G}{\partial R} - \frac{\partial \Fdep}{\partial R} \right) + \Sigma R \frac{\partial l}{\partial t} \right] \right\}.
\end{equation}
This equation (or its analogues) is most often presented without the $\partial l/\partial t$ term, e.g see \citet{Lynden1974}, \cite{pringle_accretion_1981}, \citet{Balbus1999}, \cite{rafikov_planet_2002}, and often without the deposition torque term. Dropping $\partial l/\partial t$ is certainly justified in a steady state, however, as we demonstrate in the next section, it cannot be ignored in the gap opening problem since the evolution of $\Sigma$ inevitably causes time dependence of the specific angular momentum $l$, making $\partial l/\partial t$ nonzero in general. Thus, to study the evolution of $\Sigma$ in a self-consistent way we will analyze the full equation (\ref{eq:time_evolve_sigma}).

%%%%%%%%%%%%%%%%%%%%%%%%%%%%%%%%%%%%%%%%%%%%%%%%%%
%%%%%%%%%%%%%%%%%%%%%%%%%%%%%%%%%%%%%%%%%%%%%%%%%%

\section{Initial evolution of gaps in inviscid discs}
\label{sec:initial_evolution}

%%%%%%%%%%%%%%%%%%%%%%%%%%%%%%%%%%%%%%%%%%%%%%%%%%

To study gap development using equation (\ref{eq:time_evolve_sigma}), we start by making several assumptions:
\begin{itemize}
  \item We assume the disc to be inviscid, allowing us to set $G=0$ in equation (\ref{eq:time_evolve_sigma}). The case of a viscous disc is considered in Section \ref{sec:viscosity}.
  
  \item Thermodynamic response of the disc is characterized by the locally isothermal EoS, $P = c_s^2 \Sigma$, although later on we specialize to the globally isothermal EoS.
  
  \item The action of the wave on the disc can be described through the specific angular momentum deposition function, $\fdep(R)$, such that
  \begin{equation}
  \label{eq:amfd_func}
      \frac{\partial \Fdep}{\partial R} = \Sigma\fdep(R),
  \end{equation}
\end{itemize}
where the specific form of $\fdep(R)$ depends on the physical mechanism of wave dissipation.

Furthermore, to make analytical progress we make several approximations:
\begin{itemize}
  \item We consider a linear regime, in which the relative surface density perturbation $\delta\Sigma/\Sigma_0$ is assumed to be small,
  \begin{equation}
  \Sigma(R, t) = \Sigma_0(R)\left[ 1 + \sigma(R,t)\right],~~~~~\mbox{with}~~~~~
  |\sigma(R, t)| \ll 1,
  \label{eq:linearity}
  \end{equation}
  and localized, so that $\sigma\to 0$ far from the planet. 
  
  \item The angular velocity in the disc is given by the radial force balance in the form
  \begin{equation}
      \label{eq:force_balance}
        \Omega^2 = \OmK^2 + \frac{1}{R \Sigma} \frac{\partial P}{\partial R},
  \end{equation}
  i.e. we neglect the terms $\partial v_R/\partial t$ and $v_R\partial v_R/\partial R$ in the radial component of the Navier-Stokes equation. 
  
  \item We will also assume the deviations from the Keplerian rotation to be small, so that the second term in the right-hand side of equation (\ref{eq:force_balance}) is small compared to $\OmK^2$. Nevertheless, the difference between $\Omega$ and $\OmK$ will be important when evaluating the $\partial l/\partial t$ term in equation (\ref{eq:time_evolve_sigma}).
  
  \item At the same time, in the rest of equation (\ref{eq:time_evolve_sigma}) we will set $\Omega=\OmK$.
\end{itemize}
The validity of these assumptions will be verified in Section \ref{sec:validity}.

With these assumptions and approximations we proceed as follows. Equation (\ref{eq:force_balance}) gives 
\begin{equation}
    l^2 = R^4 \Omega^2 = R^4\OmK^2 + \frac{R^3}{\Sigma} \frac{\partial P}{\partial R}.
\end{equation}
For small deviations from Keplerian rotation we can simplify
\begin{equation}
    l = R^2 \OmK \left(1 + \frac{1}{\OmK^2 R \Sigma}  \frac{\partial P}{\partial R}\right)^{1/2} \approx 
    R^2 \OmK \left(1 + \frac{1}{2} \frac{c_s^2}{\OmK^2 R}  \frac{\partial \ln P}{\partial R}\right).
    \label{eq:Om_dev}
\end{equation}
Then, using the locally isothermal EoS and the linear anzatz (\ref{eq:linearity}), we can write
\begin{align}
    \frac{\partial l}{\partial t} &\approx  
    \frac{1}{2} \frac{R c_s^2}{\OmK}  \frac{\partial \ln P}{\partial t} =\frac{1}{2} \frac{R c_s^2}{\OmK} \frac{\partial}{\partial t} \left[\frac{\partial}{\partial R}\ln\left(\Sigma_0 c_s^2\right) +
    \frac{\partial}{\partial R}\ln(1+\sigma)
    \right] \nonumber \\
    &\approx     \frac{1}{2} \frac{R c_s^2}{\OmK } \frac{\partial^2 \sigma}{\partial R \partial t}.
    \label{eq:ldot}
\end{align}
This equation demonstrates a direct relation between the time evolution of $\Sigma$, or its relative perturbation $\sigma=\delta\Sigma/\Sigma_0$, and $\dot{l} = \partial l/\partial t$. The calculation of $\dot{l}$ is the only place where we account for the deviation of $\Omega$ from $\OmK$.

Substituting equations (\ref{eq:amfd_func}), (\ref{eq:linearity}) and (\ref{eq:ldot}) into equation (\ref{eq:time_evolve_sigma}), we obtain the following equation for $\dot{\sigma} = \partial \sigma/\partial t$,
\begin{equation}    
    \Sigma_0 \dot{\sigma} - \frac{1}{R}\frac{\partial}{\partial R}\left(\frac{\Sigma_0 R c_s^2}{\OmK^2} \frac{\partial \dot{\sigma}}{\partial R} \right) = - \frac{1}{R} \frac{\partial}{\partial R} 
    \left(\frac{\Sigma_0}{\pi R \OmK}\fdep(R) \right),
    \label{eq:global_inviscid}
\end{equation}
to lowest (linear) order\footnote{Note that we replaced $\Sigma$ with $\Sigma_0$ in the last term of equation (\ref{eq:global_inviscid}), since otherwise we would be introducing a term quadratic in $\sigma$.} in $\sigma\ll 1$. 

This equation is linear in $\dot{\sigma}$ and contains only the spatial derivatives when considered as an equation for $\dot\sigma$. Therefore, if $\fdep$, $\Sigma_0$ and $c_s$ are independent of time, then $\dot\sigma$ ends up being the function of $R$ only. Thus, once we determine $\dot{\sigma}(R)$, it immediately follows that the surface density deviation $\delta\Sigma$ simply grows linearly in time,
\begin{equation}
    \frac{\delta \Sigma(R, t)}{\Sigma_0(R)} = \sigma(R,t)=\dot{\sigma}(R) \, t,
    \label{eq:lin-time}
\end{equation}
assuming that $\sigma=0$ at $t=0$. In Section \ref{sec:time_dependent} we generalize this result to the case of disc or planetary parameters varying in time, e.g. due to disc evolution or planetary migration and accretion.

Equation (\ref{eq:global_inviscid}) is a linear, diffusion-type equation for $\dot\sigma$ with the source term containing $\fdep$. If $\fdep(R)=0$ everywhere, i.e. there is no angular momentum injection across the disc, then $\sigma(R,t)=0$, as expected. But if $\fdep(R)$ is non-zero even in a finite radial interval in the disc, the diffusion-type nature of equation (\ref{eq:global_inviscid}) will make $\sigma(R,t)$ non-zero globally, see Section \ref{sec:local} and Appendix \ref{sec:full_global}. 

This is qualitatively different from the standard approach \citep{Lynden1974,pringle_accretion_1981}, in which one ignores the $\dot l$ term in equation (\ref{eq:time_evolve_sigma}). In that case one would end up with the modified version of equation (\ref{eq:global_inviscid}), without the second term in the left-hand side. As this would then be an algebraic equation for $\dot\sigma$, one would immediately obtain, using equation (\ref{eq:lin-time}),
\begin{equation}    
    \delta \Sigma(R, t) =  - \frac{t}{R} \frac{\partial}{\partial R} 
    \left(\frac{\Sigma_0}{\pi R \OmK}\fdep(R) \right).
    \label{eq:no-ldot-soln}
\end{equation}
This solution implies that $\delta\Sigma=0$ in parts of the disc where $\fdep(R)=0$, even if $\fdep$ is non-zero in other parts of the disc. This is precisely the situation illustrated in Figure \ref{fig:fig_one}, where the blue curve in panel (b) corresponding to the solution (\ref{eq:no-ldot-soln}) is zero in the co-orbital region where $\fdep(R)$ (shown in panel (a)) is also zero. Just beyond $\lsh$ from the planet where $|\fdep(R)|$ peaks, the blue curve shows two deep narrow gaps as predicted by \cite{rafikov_nonlinear_2002}. This behavior is very different from the true $\delta\Sigma$ computed by fully accounting for the $\dot l$ term, either via equation (\ref{eq:global_inviscid}; orange curve) or via the full nonlinear simulation (black curve): there is a smooth double valley structure with troughs around $|R-\Rp|\approx \lsh$ and mass is evacuated from the co-oribtal region, as also seen in other inviscid simulations \citep{cimerman_planet-driven_2021}. 

In Appendix \ref{sec:full_global} we provide a global solution of the full equation (\ref{eq:global_inviscid}) for the power-law radial profiles of the background $\Sigma_0$ and $c_s$. In the rest of this paper, when comparing the analytical predictions for gap opening based on equation (\ref{eq:global_inviscid}) with simulations, we will use that global solution, equation (\ref{eq:global_solution}). However, to gain further analytical understanding of the surface density evolution near the planet, we now invoke the local approximation.

%%%%%%%%%%%%%%%%%%%%%%%%%%%%%%%%%%%%%%%%%%%%%%%%%%
%%%%%%%%%%%%%%%%%%%%%%%%%%%%%%%%%%%%%%%%%%%%%%%%%%

\subsection{Local Approximation}
\label{sec:local}

%%%%%%%%%%%%%%%%%%%%%%%%%%%%%%%%%%%%%%%%%%%%%%%%%%

In the local (or WKB) approximation we assume that the radial scale of the gap is much smaller than $\Rp$. Then in equation (\ref{eq:global_inviscid}) we can neglect the $R$ derivatives of all variables except $\dot\sigma$ and $\fdep$, with their values being evaluated at $\Rp$. This approximation should work well for very cold and geometrically thin discs with $H/R \ll 1$, since the radial scale of $\Sigma$ variation near the gap is $\sim H$, as we show below.  Under this approximation the surface density evolution equation (\ref{eq:global_inviscid}) reduces to
\begin{equation} 
    \label{eq:local_dsigma}
    \dot{\sigma} - \Hp^2 \frac{\partial^2 \dot{\sigma}}{\partial R^2}= - \frac{1}{\pi \Rp^2 \Omp} \frac{\partial \fdep(R)}{\partial R}.
\end{equation}
This second order ODE for $\dot{\sigma}$ can be easily solved with the boundary conditions that $\sigma\to 0$ as $R\to 0$ and $R\to\infty$, yielding the following solution for $\sigma(R,t)$:  \begin{align}
\sigma(R, t) =  \, \frac{t}{2\pi}\frac{1}{\Hp^2 \Rp^2 \Omp} &\bigg[ e^{R/\Hp} \int_{0}^{R} \fdep(x) e^{ - x/\Hp} \md x 
\nonumber \\
& + e^{-R/\Hp} \int_{0}^{R} \fdep(x) e^{x/\Hp} \md x 
\nonumber \\
& -2 \sinh\left( R/\Hp \right) \int_{0}^{\infty} \fdep(x) e^{ - x/\Hp} \md x  \bigg]. 
\label{eq:full_solution}
\end{align}
In Appendix \ref{sec:full_global} we compare this local solution with the fully global solution, equation (\ref{eq:global_solution}), for the same angular momentum deposition function $\fdep(R)$ and a range of disc and planetary parameters, focusing on the deviations forced by the local approximation.

The solution (\ref{eq:full_solution}) is valid for any form of $\fdep(R)$, as long as it is sufficiently localized around $\Rp$. As a result, this solution can be applied to understand gap opening not only due to the nonlinear wave dissipation but also as a result of the linear wave damping, e.g. due to radiative losses \citep{miranda_planetary_2019,miranda_planet-disk_2020,miranda_gaps_2020}. We will now explore the solution (\ref{eq:full_solution}) further by making additional assumptions and approximations.

%%%%%%%%%%%%%%%%%%%%%%%%%%%%%%%%%%%%%%%%%%%%%%%%%%

\subsection{Globally isothermal EoS}
\label{sec:globalEoS}

%%%%%%%%%%%%%%%%%%%%%%%%%%%%%%%%%%%%%%%%%%%%%%%%%%

We now focus of the particular case of the globally isothermal EoS, with $c_s$ being independent of $R$. This EoS (like any adiabatic EoS) possesses the nice property of conserving the wave AMF $F_\mathrm{wave}$ in the absence of explicit dissipation \citep{miranda_planetary_2019,miranda_planet-disk_2020}, which considerably simplifies our analysis and interpretation of simulations; see Section \ref{sec:discussion} for the discussion of issues emerging when using the locally isothermal EoS.

Furthermore, adoption of the globally isothermal EoS allows us to make use of the results for the $\fdep(R)$ behavior obtained in \cite{cimerman_planet-driven_2021} for the globally isothermal EoS. That study, based on the theory of \cite{rafikov_nonlinear_2002}, provided a semi-analytical prescription for $\fdep(R)$, which was calibrated against simulations. For convenience, we re-iterate the detail of this particular angular momentum deposition model (correcting some errors, as necessary) in Appendix \ref{sec:deposition_models}, with a representative example of $\fdep(R)$ behavior shown in Figure \ref{fig:fig_one}a. This figure illustrates two key properties of $\fdep(R)$ in the globally isothermal case. 

First, $\fdep(R)=0$ in the coorbital region, where the planet-driven density wave has not shocked yet. Once the wave shocks at $R=\Rp\pm\lsh$, $\fdep(R)$ becomes non-zero. This behavior is typical for nonlinear damping in the barotropic case and would not occur in the case of a locally isothermal EoS, or if linear wave damping were present, see Section \ref{sec:discussion}.

Second, one can see an asymmetry in $\fdep(R)$ behavior inside and outside $\Rp$: for the adopted values of disc parameters the outer peak of $\fdep$ has somewhat larger amplitude than its inner trough. This asymmetry follows from the $\fdep(R)$ prescription in \cite{cimerman_planet-driven_2021} and gets fully accounted for by the solution (\ref{eq:full_solution}). However, to obtain further useful analytical insights, it makes sense to consider a simplified local model for $\fdep(R)$, which is symmetric relative to $\Rp$, see next.

%%%%%%%%%%%%%%%%%%%%%%%%%%%%%%%%%%%%%%%%%%%%%%%%%%

\subsection{Solution for a symmetric \texorpdfstring{$\fdep(R)$}{fdep} }
\label{sec:sym-fdep}

%%%%%%%%%%%%%%%%%%%%%%%%%%%%%%%%%%%%%%%%%%%%%%%%%%

%%%%%%%%%%%%%%%%%%%%%%%%%%%%%%%%%%%%%%%%%%%%%%%%%%
\begin{figure}
    \centering
    \includegraphics[width=0.47\textwidth]{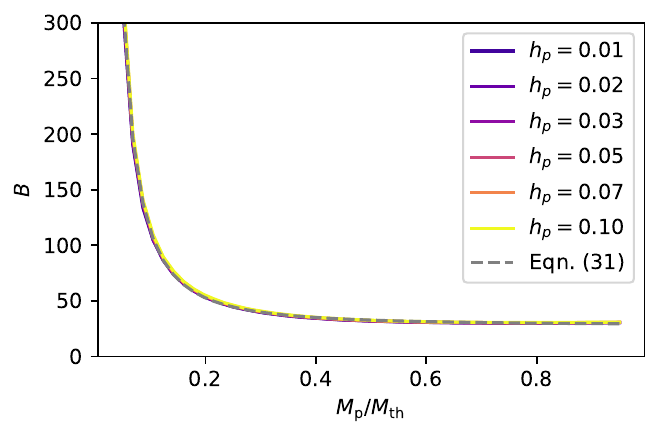}
    \caption{Dependence of the parameter $B$ defined by equation (\ref{eq:B_const}) on $\Mp/\Mth$ obtained via the procedure described in Section \ref{sec:sym-fdep} for several values of $\hp$. Note the universality of $B(\Mp/\Mth)$ as $\hp$ is varied. The gray dashed line represents the fit (\ref{eq:fit_b}). }
    \label{fig:fig_two}
\end{figure}
%%%%%%%%%%%%%%%%%%%%%%%%%%%%%%%%%%%%%%%%%%%%%%%%%%
For illustrative purposes and to obtain a simpler form of the solution (\ref{eq:full_solution}) we now neglect the aforementioned asymmetry of $|\fdep(R)|$ and assume that $\fdep(R)$ is an odd function in $R-\Rp$, bearing in mind the non-linear nature of the wave damping over a characteristic radial scale of $\lsh$:
\begin{equation}
    \fdep(R)=\frac{1}{\Sigma}\frac{\partial \Fdep}{\partial R} = \mathrm{sgn}\left(R - \Rp\right)\frac{F_{J, 0}}{\Sigmap\lsh} \varphi \left(\frac{|R - \Rp|}{\lsh} \right).
    \label{eq:fdep_sym}
\end{equation}
Here 
\begin{equation}
    F_{J, 0}  = \left( \frac{\Mp}{M_\star} \right)^2 \hp^{-3} \Sigmap \Rp^4 \Omp^2 
    \label{eq:classic_torque}
\end{equation}
is the classic scaling for the one-sided planet-driven Lindblad torque \citep{GT1980}, while $\varphi(x)$ is the dimensionless function characterising spatial pattern of angular momentum deposition, which obeys $\varphi(x)=0$ for $x<1$ (i.e. no wave dissipation prior to its shocking) and $\varphi(x)\to 0$ as $x\to \infty$. We will not specify a particular form for the symmetric $\varphi(x)$ as our key results will end up being insensitive to its detailed form. 

Using the anzatz (\ref{eq:fdep_sym}) and definition (\ref{eq:classic_torque}), and considering the local limit, i.e. $\Hp/\Rp\to 0$, $\lsh/\Rp\to 0$, the solution (\ref{eq:full_solution}) simplifies to
\begin{eqnarray}
    \sigma(R, t) = -\frac{\Omp t}{\pi}\hp\left( \frac{\Mp}{\Mth} \right)^2\left[B^{-1}\cosh\left(\frac{R - \Rp}{\Hp}\right) -\chi\left(\frac{|R - \Rp|}{\lsh}\right)\right],
    \label{eq:sym_sol_gen}
\end{eqnarray}
where the function $\eta(z)$ is defined for $z>0$ as 
\begin{eqnarray}
    \chi(z)=\left\{
    \begin{array}{ll}
    0, & z<1, \\ 
    \int_1^z\varphi(x) \cosh{\left[(z-x)\frac{\lsh}{\Hp}\right]}\, \md x, & z>1, 
    \end{array}
    \right.
    \label{eq:chi}
\end{eqnarray}
and the constant $B$ is defined via
\begin{eqnarray}
    B^{-1}=
    \int_1^\infty\varphi(x) \exp\left(-\frac{\lsh}{\Hp}x\right)\md x. 
    \label{eq:B_const}
\end{eqnarray}
One can easily verify that $\sigma\to 0$ far from the planet, for $|R - \Rp|/\lsh,|R - \Rp|/\Hp\to \infty$, as long as $\varphi(z)\to 0$ for $z\to\infty$.

In the coorbital region, within $\lsh$ from $\Rp$, the solution (\ref{eq:sym_sol_gen}) takes a particularly simple form:
\begin{eqnarray}
    \sigma(R, t) = -\frac{\Omp t}{\pi}B^{-1}\hp\left( \frac{\Mp}{\Mth} \right)^2 \cosh\left(\frac{R - \Rp}{\Hp}\right), ~~~~~|R - \Rp|<\lsh.
    \label{eq:most_simplifed_x}
\end{eqnarray}
The 'coshine' profile of this solution holds regardless of the actual structure of the angular momentum deposition function $\varphi(x)$; only the constant factor $B$ depends on the details of global $\varphi(x)$ behavior, see equation (\ref{eq:B_const}). Thus, the solution (\ref{eq:most_simplifed_x}) is a very general result for the initial shape of a gap in the coorbital region of a small planet carving it. And, very importantly, it clearly shows that $\sigma$ (or $\delta\Sigma$) is non-zero in the coorbital region, even though there is no deposition of the wave angular momentum there. This is very different from the behavior predicted by equation (\ref{eq:no-ldot-soln}).

The constant $B$ entering the solution (\ref{eq:sym_sol_gen}) depends not only on the angular momentum deposition pattern, i.e. $\varphi(z)$, but also on the ratio $\lsh / \Hp$. According to equation (\ref{eq:lsh}), $\lsh / \Hp$ is a function of $\Mp/\Mth$ and $\gamma$, therefore for a given $\varphi(x)$ and $\gamma$, one should have $B=B(\Mp/\Mth)$. One way to determine this dependence is to choose an expression for $\varphi(x)$ and evaluate the integral in (\ref{eq:B_const}) as a function of $\Mp/\Mth$. However, we have chosen a different route based on the global solution of the full equation (\ref{eq:global_inviscid}) provided by equation (\ref{eq:global_solution}) in Appendix \ref{sec:full_global} with the asymmetric $\fdep$ from Appendix \ref{sec:deposition_models}. We use this solution to extract the value of $\dot\sigma(\Rp)$ and then match it to the local, symmetric solution (\ref{eq:most_simplifed_x}) by setting $B=-\left[\pi\dot\sigma(\Rp)\right]^{-1}\Omp\hp(\Mp/\Mth)^{2}$. 

The result of this procedure is shown in Figure \ref{fig:fig_two} for a range of values of $\hp$. One can see that the results are largely independent of $\hp$, implying that the anzatz (\ref{eq:fdep_sym}) reproduces quite well the actual dependence of  $\fdep$ on $\hp$. As a result, $B$ is indeed a rather universal function of $\Mp/\Mth$ only. An offset power law in the form 
\begin{equation}
    \label{eq:fit_b}
    B(\Mp/\Mth) = 27.9 + 1.41 \left(\Mp/\Mth\right)^{-1.78}
\end{equation}
was found to provide a good fit\footnote{Using SciPy's curve\_fit  routine \citep{virtanen_scipy_2020}.} to the shape of $B(\Mp/\Mth)$. This fit was evaluated for the $\Sigma$ slope $p = 1.5$, however tests with other values of $p$ show no significant deviations. The simple analytical solution (equation \ref{eq:most_simplifed_x}) with $B$ given by equation (\ref{eq:fit_b}) will be used in various checks of our approximation, see Section \ref{sec:validity}.

%%%%%%%%%%%%%%%%%%%%%%%%%%%%%%%%%%%%%%%%%%%%%%%%%%

\subsection{Gap opening timescale}
\label{sec:time-scale}

%%%%%%%%%%%%%%%%%%%%%%%%%%%%%%%%%%%%%%%%%%%%%%%%%%

The solution (\ref{eq:most_simplifed_x}) can be re-written as
\begin{eqnarray}
    \sigma(R, t) = -\frac{t}{\tgap}\cosh\left(\frac{R - \Rp}{\Hp}\right), ~~~~~|R - \Rp|<\lsh,
    \label{eq:blerg}
\end{eqnarray}
where we have introduced a characteristic timescale 
\begin{eqnarray}
    \label{eq:time_gap}
    \tgap = \frac{P_{p}}{2 h_p} \left(\frac{\Mp}{\Mth}\right)^{-2}  B\left(\frac{\Mp}{\Mth}\right),
\end{eqnarray}
with $\Pp$ being the orbital period of the planet. The formal meaning of $\tgap$ is rather straightforward: it is the time needed for surface density at $\Rp$ to become zero, i.e. $\sigma(\Rp)\to -1$, if the solution (\ref{eq:blerg}) were to hold at all times. 

Obviously, other parts of the the gap profile would reach $\Sigma=0$ even earlier than at $\Rp$, for instance, at $R=\Rp\pm\lsh$ this would happen at $t=\tgap/\cosh(\lsh/\Hp)<\tgap$. And, of course, the solution (\ref{eq:blerg}) would fail before $\sigma$ could reach $-1$ simply because it was obtained in the linear limit, something that we will demonstrate in Section \ref{sec:validity}. Nevertheless, $\tgap$ can still be considered as an order of magnitude estimate (or an upper limit) of the timescale of gap opening in inviscid discs. For example, for a $30 M_{\earth}$ planet at $\Rp=50$ AU in a $\hp = 0.07$ disc around a $1 M_{\sun}$ star ($\Mp/\Mth \approx 0.26$) equations (\ref{eq:fit_b}) \& (\ref{eq:time_gap}) give $\tgap\approx 1.6$ Myr.

The expression (\ref{eq:time_gap}) for $\tgap$ is rather similar to the gap-opening time estimate in \citet[][equation (26)]{rafikov_planet_2002}, although instead of $B$ that paper has a factor of $\sim(\lsh/\Hp)^2$ (as their estimate did not use a particular model of the wave dissipation). However, the overall qualitative behavior of the two timescales as a function of $\Mp$ and $\hp$ is very similar.

%%%%%%%%%%%%%%%%%%%%%%%%%%%%%%%%%%%%%%%%%%%%%%%%%%
\begin{figure}
    \centering
    \includegraphics[width=0.48\textwidth]{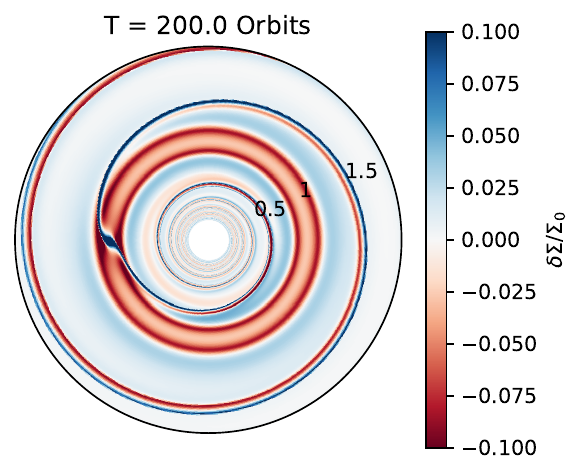}
    \caption{Map of the relative surface density perturbation $\sigma=\delta\Sigma/\Sigma$ at $t=200\Pp$ from an inviscid simulation with $\Mp/\Mth= 0.25, \hp = 0.05,$ and $ p = 1.5.$. 
    It illustrates the planet-driven spiral density waves in the inner and outer discs, the double gap structure around the planetary orbit, and the additional inner gap at $R \approx 0.5 R_p$ appearing due to the formation and non-linear evolution of the secondary arm in the inner disc.
    An animated version of this figure including the non-linear period of surface density evolution and onset of the Rossby wave instability is available in the supplemental materials.}
    \label{fig:2d_visual}
\end{figure}
%%%%%%%%%%%%%%%%%%%%%%%%%%%%%%%%%%%%%%%%%%%%%%%%%%

%%%%%%%%%%%%%%%%%%%%%%%%%%%%%%%%%%%%%%%%%%%%%%%%%%
\begin{figure}
    \centering    
    \includegraphics[width=0.48\textwidth]{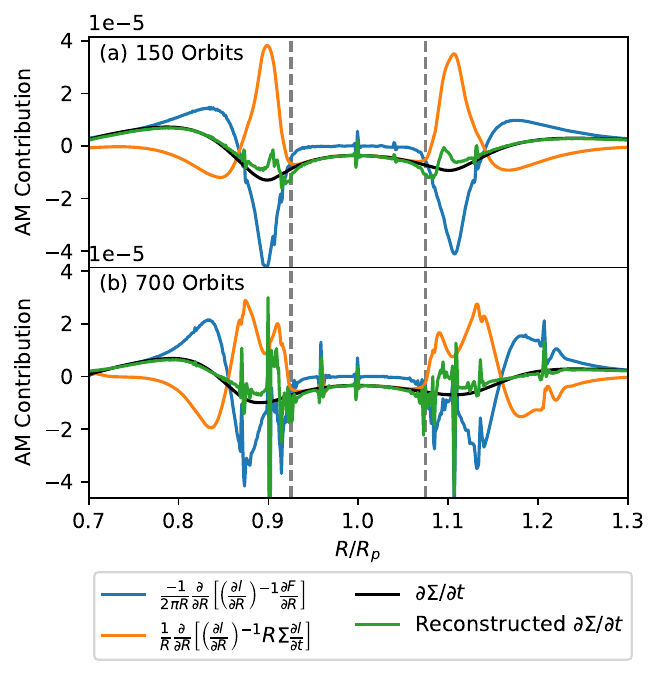}
    \caption{
    Different contributions to the angular momentum balance in equation (\ref{eq:time_evolve_sigma}) as measured in a simulation with $\Mp = 0.25 \Mth, \hp=0.05, p=1.5$ at $t = 150 \Pp$ (top) and $t = 700 \Pp$ (bottom). We show $\partial \Sigma/\partial t$ (red) and the azimuthally averaged angular momentum deposition term (blue) and $\dot{l}$ term (orange); green curve ('reconstructed $\partial \Sigma/\partial t$') is the sum of the latter two and would match the red curve in the absence of numerical artefacts. Dashed gray lines are at $|R-\Rp|=\lsh$ away from the planet on each side. In both panels, the $\partial l/\partial t$ contribution is very important and is at least as large as the angular momentum deposition contribution; both are considerably larger than $\partial \Sigma/\partial t$. Note that 
    for this simulation $\tgap\approx 7500\Pp$ and $t_\mathrm{nl}\approx 520\Pp$ (see Equation (\ref{eq:tnl_fixed})), so that panel (b) is in the (mildly) nonlinear regime.}
    \label{fig:split_up_sigma}
\end{figure}
%%%%%%%%%%%%%%%%%%%%%%%%%%%%%%%%%%%%%%%%%%%%%%%%%%

%%%%%%%%%%%%%%%%%%%%%%%%%%%%%%%%%%%%%%%%%%%%%%%%%%
\begin{figure*}
    \centering
    \includegraphics[width=\textwidth]{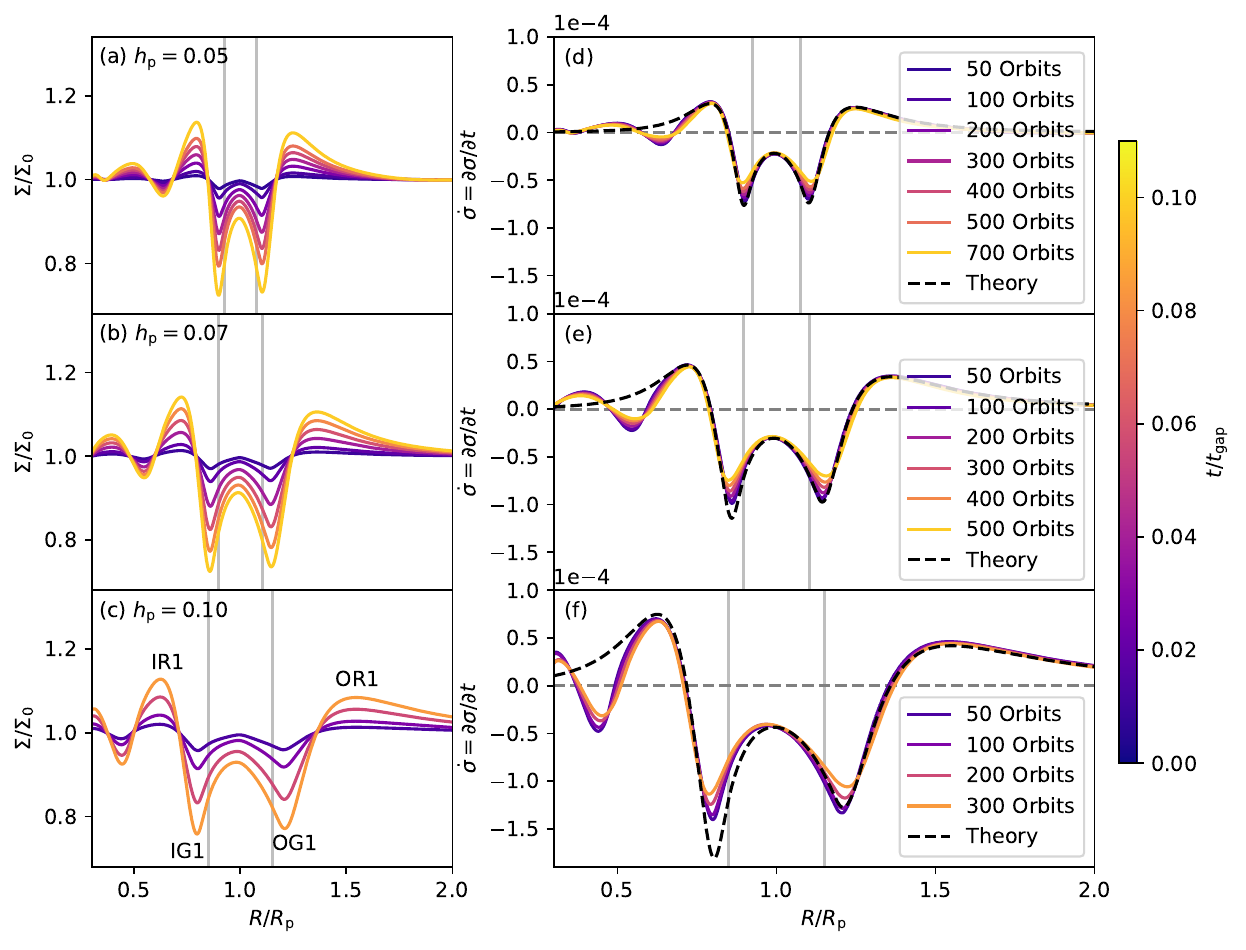}
    \caption{Evolution of the azimuthally-averaged surface density measured in simulations with a $\Mp = 0.25 \Mth$ planet in globally-isothermal discs with $p=1.5$ and (from top to bottom) $\hp = 0.05, 0.07, 0.1$.  
    Vertical gray lines show positions $\pm l_{sh}$ away from $\Rp$.
    Left: Azimuthally averaged $\Sigma/\Sigma_0$ at different moments of time, universally exhibiting a double-valley profile \citep{rafikov_planet_2002}. Line colours are mapped to time (shown in right panels) normalised to $\tgap$, see equation (\ref{eq:time_gap}), to highlight that gap behaviour opening is a function of this parameter. Panel (c) also includes our labelling of the different parts of the central gap structure as used in Figure \protect \ref{fig:gas_spacings}.
    Right: Time derivative of $\delta\Sigma/\Sigma$, i.e. $\dot\sigma$ at the same moments of time as in the left panels (indicated in the legend). Solid lines are the simulation results, while the black dashed curves are the global solutions (\ref{eq:global_solution}) for $\dot\sigma$ based on our linear theory. This figure illustrates the self-similar nature of the initial gap development, with $\sigma(R,t)$ linearly growing in time, while maintaining a fixed profile in $R$. Note the discrepancy between the theory and simulations at $R\approx 0.4\Rp$ (bottom) to $0.7\Rp$ (top), where the inner gap appears because of the secondary arm formation. See text for details.}
    \label{fig:compare_same_mass}
\end{figure*}
%%%%%%%%%%%%%%%%%%%%%%%%%%%%%%%%%%%%%%%%%%%%%%%%%%

%%%%%%%%%%%%%%%%%%%%%%%%%%%%%%%%%%%%%%%%%%%%%%%%%%
\begin{figure*}
    \centering
    \includegraphics[width=0.95\textwidth]{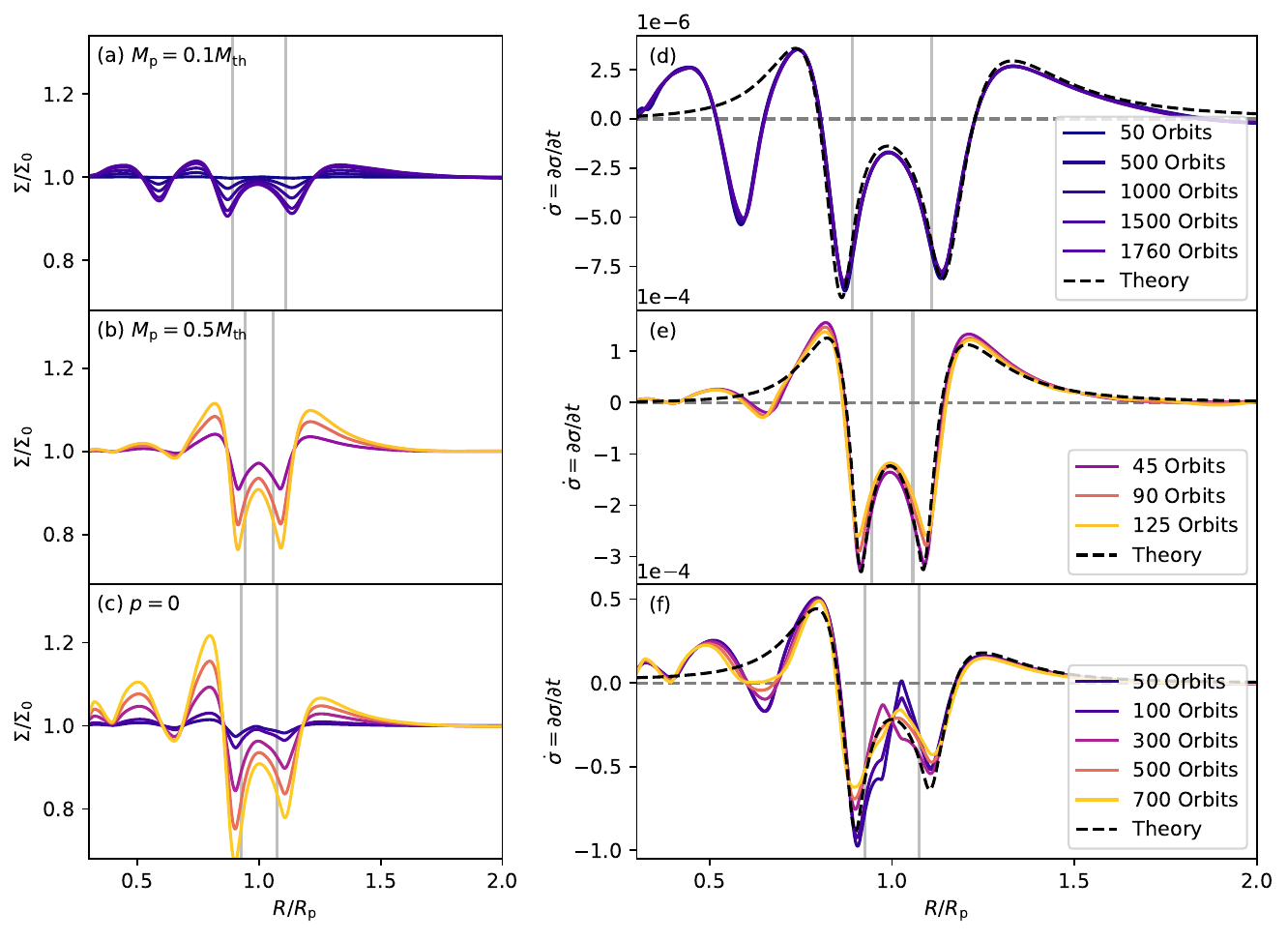}
    \caption{Same as Figure \ref{fig:compare_same_mass}, now illustrating how well the global linear solution (\ref{eq:global_solution}) works for planets of mass $\Mp = 0.1 \Mth$ and $\Mp = 0.5 \Mth$ in a $\hp = 0.05, p = 1.5$ disc (upper two rows), and for a $\Mp = 0.25 \Mth$ planet in a $\hp = 0.05, p = 0$ disc (bottom row).
    The colours for each line are mapped to $t/\tgap$, as in Figure \ref{fig:compare_same_mass}. See text for details.}
%    As expected the low $\Mp$ case evolves linearly for a far longer period than the high $\Mp$ case as $\partial \sigma/\partial t$ remains fixed, but for all cases the initial evolution is well represented by our model. For the $p = 0$ case there are additional short term deviations in the evolution track. In the low mass case because the evolution is so slow the time derivatives are equal over an order of magnitude in time.}
    \label{fig:compare_diff_mass}
\end{figure*}
%%%%%%%%%%%%%%%%%%%%%%%%%%%%%%%%%%%%%%%%%%%%%%%%%%

%%%%%%%%%%%%%%%%%%%%%%%%%%%%%%%%%%%%%%%%%%%%%%%%%%
%%%%%%%%%%%%%%%%%%%%%%%%%%%%%%%%%%%%%%%%%%%%%%%%%%

\section{Invisicd simulations}
\label{sec:results}

%%%%%%%%%%%%%%%%%%%%%%%%%%%%%%%%%%%%%%%%%%%%%%%%%%

We now test our analytical results for initial gap evolution obtained in the previous section, using fully non-linear inviscid hydrodynamical 2D simulations with Athena++ \citep{stone_athena_2020} obtained as a part of \cite{cimerman_planet-driven_2021}. Full details of the numerical setup can be found in Appendix \ref{sec:numerical_a}.

We show a typical 2D snapshot from a globally isothermal simulation with $\hp=0.05$ at $t=200\Pp$ in Figure \ref{fig:2d_visual} to remind the reader of the main 2D features of the disc-planet interaction. One can see a prominent one-armed spiral density wave launched by a $\Mp=0.25$ planet in the inner and outer parts of the disc, which is time-independent in the frame co-rotating with the planet (after about $10\Pp$ from the introduction of the planetary potential). Nonlinear evolution of the wave leads to its shocking after travelling a radial distance $\lsh$ from the planet, injection of the wave angular momentum into the disc and formation of a double-valley gap around $\Rp$, which is deepest around $|R-\Rp|\approx \lsh$ and is shallower in the co-orbital region. Formation of this gap has been studied analytically in Section \ref{sec:initial_evolution}, and in Section \ref{sec:theorycomparison} we provide a comparison with simulations. Over time, this gap deepens while maintaining the same overall radial profile until the onset of the Rossby Wave Instability (RWI), beyond which point the evolution can no-longer be treated as approximately axisymmetric.

In the inner disc, at $R \approx 0.5 \Rp$, an additional inner gap forms, as a result of (linear) emergence and (nonlinear) dissipation of the secondary spiral arm in the inner disc \citep{Dong2017,bae_formation_2017,miranda_multiple_2019}. The inner gap is also the reason for the depression in the black curve (showing the results of $\hp=0.03$ simulation) around $0.65\Rp$ in Figure \ref{fig:fig_one}b.

%%%%%%%%%%%%%%%%%%%%%%%%%%%%%%%%%%%%%%%%%%%%%%%%%%

\subsection{The relative importance of the \texorpdfstring{$\dot l$}{l dot} term}
\label{sec:orderofmaginitue}

%%%%%%%%%%%%%%%%%%%%%%%%%%%%%%%%%%%%%%%%%%%%%%%%%%

Our analysis in Section \ref{sec:initial_evolution} uncovered the importance of the $\dot l$ term in equation (\ref{eq:time_evolve_sigma}) for proper understanding of surface density evolution in inviscid discs. We now illustrate this fact directly by plotting in Figure \ref{fig:split_up_sigma} the different contributions to the angular momentum budget --- individual terms (with viscosity set to zero) in equation (\ref{eq:time_evolve_sigma}) --- evaluated using the data from a simulation at two different times. We can make several observations.

First and foremost, the amplitude of the $\dot l$ term (orange curve) is large and is close to the magnitude of the angular momentum deposition term (blue curve). They both dominate the angular momentum budget even at late time, see panel (b). Second, the amplitude of the $\partial\Sigma/\partial t$ term (red curve) is considerably lower than that of the other two terms. This means that $\dot l$ and $\Fdep$ terms nearly cancel each other to result in a small net $\partial\Sigma/\partial t$. This near cancellation also naturally explains the low amplitude in Figure \ref{fig:fig_one}b of $\delta\Sigma/\Sigma$ computed with the $\dot l$ term or taken from simulations compared to $\delta\Sigma/\Sigma$ computed without the $\dot l$ term (blue). Third, in the planetary co-orbital region (between the vertical dashed lines) $\partial \Fdep/\partial R$ is almost uniformly zero, as expected \citep{goodman_planetary_2001,rafikov_nonlinear_2002}. However, $\Sigma$ still evolves in this region, see e.g. black curve in Figure \ref{fig:fig_one}b. Figure \ref{fig:split_up_sigma} clearly shows that this evolution is driven solely by the $\dot l$ term, as the orange and red curves overlap in the coorbital region. This means, in particular, that when $l$ evolves as a result of time-varying radial pressure gradient,  a one-to-one relationship between $\partial\Fdep/\partial R$ and radial mass flux in the disc does not exist. Fourth, the green curve in Figure \ref{fig:split_up_sigma} shows the sum of the $\dot l$ and $\Fdep$ terms, which in theory should equal $\partial\Sigma/\partial t$ (red curve). This is indeed the case except at radii where the $\dot l$ term peaks, which we attribute to numerical artefacts.

In retrospect, the importance of $\dot l$ term could have been gleaned upon even without this comparison against simulations. Indeed, let us first neglect this term in equation (\ref{eq:time_evolve_sigma}). Then we can estimate in the local limit, to an order of magnitude, 
\begin{equation}
\label{eq:15}
    \frac{\partial \Sigma}{\partial t} = -\frac{1}{2 \pi R} \frac{\partial }{\partial R}\left[\left(\frac{\partial l}{\partial R}\right)^{-1}\frac{\partial \Fdep}{\partial R} \right] \sim \frac{1}{\Rp^2 \OmK} \frac{\partial^2 \Fdep}{\partial R^2} \sim \frac{F_{J,0}}{\Rp^2 \OmK \lsh^2 },
\end{equation}
where we assumed $\partial^2 \Fdep/\partial R^2\sim F_{J,0}/\lsh^2$. But now let us evaluate the magnitude of the neglected $\dot l$ term using equation (\ref{eq:ldot}) and the result (\ref{eq:15}):
\begin{equation}
    \Sigma R \frac{\partial l}{\partial t}  
    \sim \frac{\Rp^2 c_s^2}{\OmK} \frac{\partial^2 \Sigma}{\partial R \partial t}\sim \frac{\Hp^2}{\lsh^2} \frac{F_{J,0}}{ \lsh }.
\end{equation}
For a $\Mp\sim\Mth$ planet $\lsh \sim \Hp$ and one obtains $\Sigma R (\partial l/\partial t) \sim F_{J,0}/\lsh \sim \partial \Fdep/ \partial R$. Thus, for gap opening in inviscid discs the $\dot l$ term is of equal importance to the angular momentum deposition term,  just as we observed in Figure \ref{fig:split_up_sigma}. This term is significant even when the value of $\Omega$ is well approximated by $\Omega_\mathrm{K}$.

%%%%%%%%%%%%%%%%%%%%%%%%%%%%%%%%%%%%%%%%%%%%%%%%%%

\subsection{Comparison to analytic solutions}
\label{sec:theorycomparison}

%%%%%%%%%%%%%%%%%%%%%%%%%%%%%%%%%%%%%%%%%%%%%%%%%%

%%%%%%%%%%%%%%%%%%%%%%%%%%%%%%%%%%%%%%%%%%%%%%%%%%
\begin{figure}
    \centering
    \includegraphics[width=0.48\textwidth]{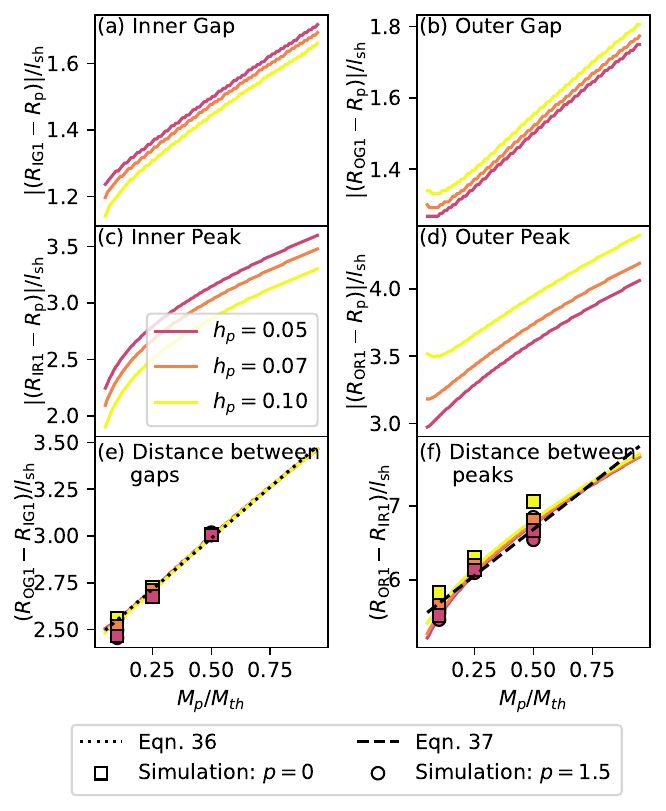}
    \caption{
    Locations of the main peaks and troughs around $\Rp$ {\bf (scaled by $1/\lsh$)} determined using the semi-analytic solution (\ref{eq:global_solution}) (curves of different color for different $\hp$) and 
    simulations (circles/squares). Locations are labelled according to the convention of \citet{dong_multiple_2018} and illustrated in Figure \ref{fig:compare_same_mass}. We find that the distances between the main troughs and the main peaks in a disc are dependent on $h_p$ only through $l_{sh}$, see panels (e) and (f) where the dashed lines show the fits (\ref{eq:gaps-fit})-(\ref{eq:extrema}). 
    }
    \label{fig:gas_spacings}
\end{figure}
%%%%%%%%%%%%%%%%%%%%%%%%%%%%%%%%%%%%%%%%%%%%%%%%%%

Having shown the importance of the $\dot l$ term, we can now compare our analytic solutions for gap opening with simulations. Left columns of Figures \ref{fig:compare_same_mass} and \ref{fig:compare_diff_mass} show $\Sigma$ profiles in the vicinity of the planet at different moments of time, taken from our simulations with different planet and disc parameters (indicated in panels) to emphasize the broad applicability of our solutions. In these figures we exclude $\Sigma$ measurements after the activation of the RWI; the typical maximum deviation of $\Sigma$ from its initial value is $\approx 20\%$.

In all our simulation and at all times, the radial profiles of $\Sigma$ evolve maintaining the same overall general shape of a double-valley gap. There is a clear but slow evacuation of mass right near the planet, at $\Rp$, leaving a ribbon of gas in the coorbital region surrounded by the two deep troughs/gaps, at $R=R_\mathrm{OG1}$ and $R_\mathrm{IG1}$ \citep[using the labelling convention of][]{dong_multiple_2018}, on each side of the planet just over $\lsh$ away, see Figure \ref{fig:compare_same_mass}c. Just outside the troughs the density peaks at $R_\mathrm{OR1}$ and $R_\mathrm{IR1}$ (outer and inner peaks/rings, respectively) as the mass gets expelled from the main gap region. This radial pattern in robustly reproduced in all our calculations. In the inner disc one can also see other, weaker overdensities and gaps; for $\hp = 0.05$ these gaps occur at $R \approx 0.6 \Rp \text{ and } 0.3 \Rp$ as a result of secondary and tertiary arms forming and shocking in such low $\hp$ discs \citep{bae_formation_2017,miranda_multiple_2019}. 

In Figure \ref{fig:gas_spacings} we show the locations of the main outer and inner troughs and peaks, as a function of $\Mp$ for different $\hp$. These locations were determined using our global solution (\ref{eq:global_solution}). One can see that even when normalised by $\lsh$, the separations of these features from $\Rp$ still show some $\hp$-dependence, however, it vanishes when one considers the distance between the troughs (gaps) $R_\mathrm{OG1}-R_\mathrm{IG1}$ and peaks (rings) $R_\mathrm{OR1}-R_\mathrm{IR1}$, see panels (e), (f). The dependence of these distances on $\Mp$ can be fitted as
\begin{align}
|R_\mathrm{OG1} - R_\mathrm{IG1}| &\approx \left(1.09 \frac{\Mp}{\Mth} + 2.44 \right)\lsh \,  ,
\label{eq:gaps-fit}\\
|R_\mathrm{OR1} - R_\mathrm{IR1}| &\approx \left(2.50 \frac{\Mp}{\Mth} + 5.44 \right)\lsh.
\label{eq:extrema}
\end{align}
Note that as $\lsh$ depends on $M_p/M_{th}$, these equations do not imply a linear relationship of these separations with $\Mp$.
 
%\begin{align}
%|R_\mathrm{OG1} - R_\mathrm{IG1}| &\approx \Hp \left(0.94 \frac{\Mp}{\Mth} + 2.10 \right)
%\left(\frac{\Mp}{\Mth}\right)^{-2/5} \, ,
%\label{eq:gaps-fit}\\
%|R_\mathrm{OR1} - R_\mathrm{IR1}| &\approx  \Hp
%\left(2.15 \frac{\Mp}{\Mth} + 4.68 \right) %\left(\frac{\Mp}{\Mth}\right)^{-2/5},
%\label{eq:extrema}
%\end{align}

% \begin{align}
%|R_\mathrm{OG1} - R_\mathrm{IG1}| &\approx 0.86 \Hp \left(1.09 \frac{\Mp}{\Mth} + 2.44 \right)
%\left(\frac{\Mp}{\Mth}\right)^{-2/5},
%\,  ,
%\label{eq:gaps-fit}\\
%|R_\mathrm{OR1} - R_\mathrm{IR1}| &\approx 
%0.86 \Hp
%\left(2.50 \frac{\Mp}{\Mth} + 5.44 \right) \left(\frac{\Mp}{\Mth}\right)^{-2/5}
% \label{eq:extrema}
%\end{align}

Analytical solutions for gap opening derived in Section \ref{sec:initial_evolution} and Appendix \ref{sec:full_global} uniformly predict that $\sigma=\delta \Sigma/\Sigma_0$ grows linearly in time, so that $\partial \sigma/\partial t$ must be a function of $R$ only, see equations (\ref{eq:full_solution}), (\ref{eq:sym_sol_gen}), (\ref{eq:global_solution}). This is indeed what we find, as illustrated in the right columns of Figures \ref{fig:compare_same_mass} and \ref{fig:compare_diff_mass}, where we display $\partial \sigma/\partial t$ at the same moments of time as in the left panels. One can see that self-similarity of $\sigma$ works remarkably well, with $\dot\sigma$ profiles at different times falling essentially on top of each other over many orbits. For example, in the low mass case (Figure \ref{fig:compare_diff_mass}a,d) $\dot{\sigma}$ is almost entirely unchanged across many orders of magnitude in time. For this planetary mass, $\tgap$ is very long and $t/\tgap$ keeps $\sigma$ in the linear regime for very long time (this is why all curves there are blue, according to our color scheme). 

The black dashed line in the right panels shows the global version of the gap opening solution (\ref{eq:global_solution}), computed for $\fdep$ specified in Appendix \ref{sec:deposition_models}. One can see that in most cases it reproduces the actual $\dot\sigma$ from simulations remarkably well, given the approximations that went into deriving it. The analytical solution always fails at reproducing the additional gaps in the inner disc, as it was not designed to account for them in the first place due to the nature of the angular momentum deposition function used in \citet{cimerman_planet-driven_2021}; in the outer disc it works very well. The solution (\ref{eq:global_solution}) also works well at predicting the relative depths of the main troughs near the planet, as well as the overall evolution in the coorbital region and at $R\sim \Rp\pm\lsh$. The agreement gets slightly worse in hotter discs, see \ref{fig:compare_same_mass}f, but it is still reasonably good. Even for higher $\Mp=0.5\Mth$ (Figure \ref{fig:compare_diff_mass}b,e), when disc-planet coupling starts approaching the nonlinear regime, the solution (\ref{eq:global_solution}) is still fully reliable. 

As $t$ increases, $\partial\sigma/\partial t$ start to slowly deviate (become shallower) from their universal, early-time shape, which is most easily seen in Figure \ref{fig:compare_same_mass}. This is caused by the eventual breakdown of the assumptions used in deriving the linear solutions in Section \ref{sec:initial_evolution}, mainly that of $\sigma\ll 1$. In Appendix \ref{sec:validity} we use the results of our simulations, which fully incorporate all nonlinear effects, to carefully assess the range of validity of our linear solutions. We draw three main conclusions from this exercise.

First, the linear model provides a good description (with relative accuracy of $10\%$) of the planet-driven gap evolution for gap depths 
\begin{eqnarray}
|\delta\Sigma|/ \Sigma_0 < 0.2, 
\label{eq:sig_max}
\end{eqnarray}
regardless of planet/simulation parameters. Second, the time it takes for our linear solutions to deviate by $\sim 10\%$ from the results of simulations is, quite universally,
\begin{eqnarray}
    \label{eq:tnl_fixed}
    t_\mathrm{nl} \approx 0.07 \,\tgap. 
\end{eqnarray}
Third, the nonlinearity of $\sigma=\delta\Sigma/\Sigma$ is more important for driving these deviations than the departure of $\Omega$ from $\OmK$ (except in equation (\ref{eq:Om_dev})).

To summarize, our linear gap opening model explicitly accounting for $\dot l$ term correctly predicts the initial shape of the gap and its time evolution in the inviscid case for a range of planetary masses and disc properties, as long as the conditions (\ref{eq:sig_max}) \& (\ref{eq:tnl_fixed}) are observed.

%%%%%%%%%%%%%%%%%%%%%%%%%%%%%%%%%%%%%%%%%%%%%%%%%%
%%%%%%%%%%%%%%%%%%%%%%%%%%%%%%%%%%%%%%%%%%%%%%%%%%

\section{Viscous Discs} 
\label{sec:viscosity}

%%%%%%%%%%%%%%%%%%%%%%%%%%%%%%%%%%%%%%%%%%%%%%%%%%

Our results so far were obtained for a purely inviscid disc, and it is natural to ask how they would change  when the disc has a non-zero viscosity $\nu$. Given the tendency of viscosity to smooth out any density inhomogeneities, we expect that it may become important when substantial gradients of $\Sigma$ develop in the process of gap formation.

To verify this expectation we first consider the evolution equation (\ref{eq:time_evolve_sigma}) with the viscous term included, and derive a generalization of the linear equation (\ref{eq:global_inviscid}). Using the anzatz (\ref{eq:linearity}) and assuming that $\Omega\approx \OmK$, we can write the gradient of the viscous angular momentum flux as 
\begin{align}
    \frac{\partial G}{\partial R} &= \frac{\partial}{\partial R}\left( -2 \pi R^3 \nu \Sigma \frac{\partial \Omega}{\partial R}\right) \approx \frac{\partial G_0}{\partial R} + \frac{\partial \delta G}{\partial R},
\end{align}
where $G_0 = 3 \pi \nu \Sigma_0 l$ and $\delta G = G_0 \sigma$. 

We will also assume that prior to the introduction of the planet the disc is in a viscous steady state, as we are interested only in the planet-driven evolution. Then $G_0 = \dot{M} l$, where $\dot M$ is the spatially constant mass accretion rate, and the $\partial G_0/\partial R$ term ends up providing no contribution in equation (\ref{eq:time_evolve_sigma}). As a result, equation (\ref{eq:global_inviscid}) generalizes to
\begin{align}
    \Sigma_0 \frac{\partial \sigma}{\partial t} =& \frac{1}{R} \frac{\partial}{\partial R} 
        \left[\frac{1}{\pi R \Omega_K} \frac{\partial}{\partial R}\left(G_0\sigma\right) \right] - \frac{1}{R} \frac{\partial}{\partial R} 
    \left[\frac{\Sigma_0}{\pi R \Omega_K}\fdep(R) \right] \nonumber \\
    &+ \frac{1}{R}\frac{\partial}{\partial R}\left[\frac{\Sigma_0 R c_s^2}{\OmK^2} \frac{\partial^2 \sigma}{\partial R\partial t} \right].
\end{align}
Under the local approximation (see Section \ref{sec:local}) it reduces to
\begin{equation} 
    \label{eq:local_visc_dsigma}
    \frac{\partial \sigma}{\partial t} = 3 \nu \frac{\partial^2 \sigma}{\partial R^2} - \frac{1}{\pi \Rp^2 \Omp} \frac{\partial \fdep(R)}{\partial R} + \Hp^2 \frac{\partial^3 \sigma}{\partial R^2\partial t},
\end{equation}
where $\nu$ is evaluated at $\Rp$. This equation is not as amenable to a time dependent soluton as the inviscid case\footnote{It is possible, though not very informative for our present purposes, to solve it using a Laplace transform.}. Nonetheless we can still analyze it to assess the effect of viscosity compared to the $\dot l$ term, which we do in Section \ref{sec:viscous_analytic}, subsequently comparing our estimates to viscous simulations in Section \ref{sec:visc_compare}.

One can also easily find a steady state solution of the equation (\ref{eq:local_visc_dsigma}) by setting $\partial / \partial t = 0$ in it:
\begin{align}
%    \frac{\partial^2 \sigma}{\partial R^2} &= \frac{1}{3 \nu} \frac{1}{\pi R_p^2 \Omega_p} \frac{\partial \fdep}{\partial R} \\ 
    \sigma(R) &= \frac{1}{3 \alpha H^2 \pi R_p^2 \Omega_p^2} \int_0^R \fdep(x) dx \,.
    \label{eq:vss_solution}
\end{align}
Then,  given a model for $\fdep(R)$, one directly obtains the surface density of a viscous steady state.  
For low-mass planets embedded in adiabatic, low viscosity discs, for which the assumption of angular momentum deposition by weakly non-linear shocking applies, Equation (\ref{eq:vss_solution}) implies that in the viscous steady state $\sigma$ is constant in the gap region spanning radially between $\Rp - \lsh$ and $\Rp + \lsh$.

We also note that the solution (\ref{eq:vss_solution}) has been derived here by assuming shallow gaps,  although it has also been obtained under more general conditions, see e.g.  \cite{dempsey_pileups_2020}.   We will compare this solution to some known results on gap depth  \citep[e.g.][]{kanagawa_formation_2015, kanagawa_mass_2015, duffell_empirically_2020} in Section  \ref{sec:compare}.

%%%%%%%%%%%%%%%%%%%%%%%%%%%%%%%%%%%%%%%%%%%%%%%%%%

\subsection{Relative importance of \texorpdfstring{$\dot{l}$}{l dot} and viscous effects}
\label{sec:viscous_analytic}

%%%%%%%%%%%%%%%%%%%%%%%%%%%%%%%%%%%%%%%%%%%%%%%%%%

As we will show, viscosity does not play a major role in the very early stages of gap opening, but eventually, after some time $t_{l-\nu}$ the viscous term starts to dominate over the $\dot l$ term and to govern the subsequent gap development. This expectation is also supported by our simulations in Section \ref{sec:visc_compare}. 

With this in mind, we can compare the roles of the viscous and $\dot l$ term in the angular momentum balance at early times by using our linear solution in the general form $\sigma(R,t)=t\dot\sigma(R)$, see equation (\ref{eq:lin-time}). Using for simplicity the (local) equation (\ref{eq:local_visc_dsigma}), one finds
\begin{equation}
    \frac{\text{viscous term}}{\dot l~\text{term}} =\frac{3 \nu \left(\partial^2 \sigma/\partial R^2\right)}{\Hp^2 \left(\partial^3 \sigma/\partial R^2\partial t\right)}=\frac{3 \nu t \left(\partial^2 \dot\sigma/\partial R^2\right)}{\Hp^2 \left(\partial^2 \dot\sigma/\partial R^2\right)}=3 \alpha \Omp t,
    \label{eq:terms_rat}
\end{equation}
where we employed the $\alpha$-viscosity anzatz (\ref{eq:nu}) with $\nu=\alpha\Hp^2\Omp$. Also, \citet{miranda_gaps_2020} have shown that in discs with $\alpha \lesssim 10^{-2}$ viscosity has a minimal effect on damping of the planet-driven density waves. This means that $\fdep$ term in equation (\ref{eq:local_visc_dsigma}) would remain essentially the same even in viscous discs. Equation (\ref{eq:terms_rat}) then implies that viscosity becomes the dominant component of evolution when 
\begin{align}
    t\gtrsim t_{l-\nu}~~~~\mbox{with}~~~~t_{l-\nu} = \frac{1}{3 \alpha \Omp} = \frac{\Pp}{6 \pi \alpha}.
    \label{eq:viscous_time-scale}
\end{align}
Prior to $t$ reaching $t_{l-\nu}$ the linear solutions obtained in Section \ref{sec:initial_evolution} should work fine. 

We note that in a self-luminous, purely viscously heated disc $t_{l-\nu}$ would be equal (up to a constant factor) to the thermal (or cooling) timescale $t_\mathrm{c}\sim \alpha^{-1}\Omega^{-1}$ \citep{Gammie2001}. However, protoplanetary discs are passively heated by their central stars, so that their thermal timescale is different from (\ref{eq:viscous_time-scale}). Also, for sub-$\Mth$ planets $t_{l-\nu}$ is shorter than the time $\alpha^{-1}\Omega^{-1}(\lsh/\Hp)^2 \sim \alpha^{-1}\Omega^{-1}(\Mp/\Mth)^{-4/5}$ it would take for viscosity to fill the gap of width $\sim \lsh$; it is also much shorter than the local viscous time of the disc $t_\nu\sim\alpha^{-1}\Omega^{-1}\hp^{-2}$.

%%%%%%%%%%%%%%%%%%%%%%%%%%%%%%%%%%%%%%%%%%%%%%%%%%
\begin{figure}
    \centering    
    \includegraphics[width=0.48\textwidth]{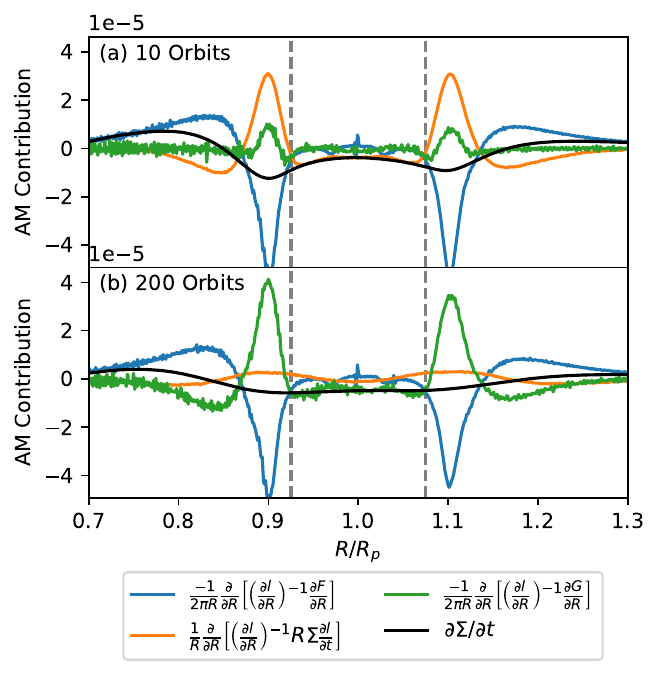}
    \caption{Different contributions to the angular momentum balance in equation (\ref{eq:time_evolve_sigma}), similar to Figure \ref{fig:split_up_sigma}, but now with the viscous term added (green curve). Data are from our $\Mp = 0.25 \Mth, \hp=0.05, p=1.5, \alpha=10^{-3.5}$ simulation at $t = 10 \Pp$ and $200 \Pp$.
    Note the change of the relative amplitudes of the $\dot l$ (orange) and viscous (green) terms as $t$ increases, and the steady decay of the $\dot l$ term compared to the $\fdep$ term (blue). See text for details.
    }
    \label{fig:split_up_sigma_viscous}
\end{figure}
%%%%%%%%%%%%%%%%%%%%%%%%%%%%%%%%%%%%%%%%%%%%%%%%%%

%%%%%%%%%%%%%%%%%%%%%%%%%%%%%%%%%%%%%%%%%%%%%%%%%%
\begin{figure}
    \centering
    \includegraphics[width=0.48\textwidth]{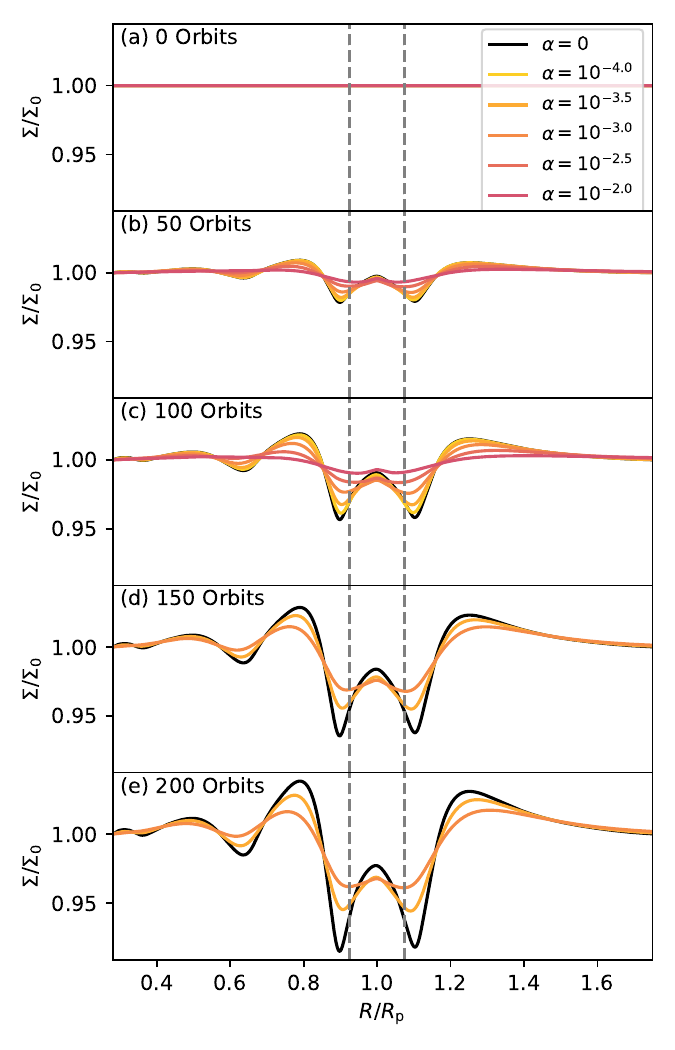}
    \caption{
    Evolution of the gap profile as a function of time (different panels) for different values of $\alpha$-viscosity (curves of different color). Data are from simulations with $\Mp/\Mth = 0.25, p=1.5, \hp=0.05$. Dashed lines show $R=\Rp\pm\lsh$. While in the inviscid case the gap depth grows linearly with $t$ in a self-similar fashion, in the viscous case the shape of the gap changes over time, leading to a shallower and smoother bottom. Note that not all simulations were run for the same length of time (some values of $\alpha$ are missing in the lower panels).
    }
    \label{fig:viscous_figure_1}
\end{figure}
%%%%%%%%%%%%%%%%%%%%%%%%%%%%%%%%%%%%%%%%%%%%%%%%%%
%%%%%%%%%%%%%%%%%%%%%%%%%%%%%%%%%%%%%%%%%%%%%%%%%%
\begin{figure}
    \centering
    \includegraphics[width=0.48\textwidth]{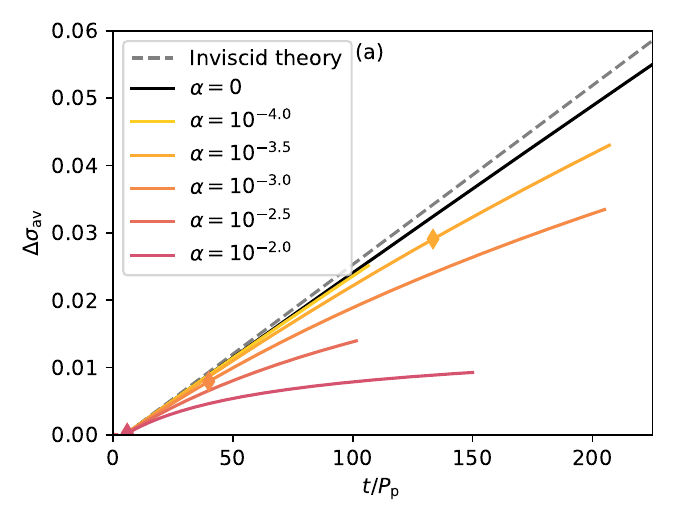}
    \includegraphics[width=0.45\textwidth]{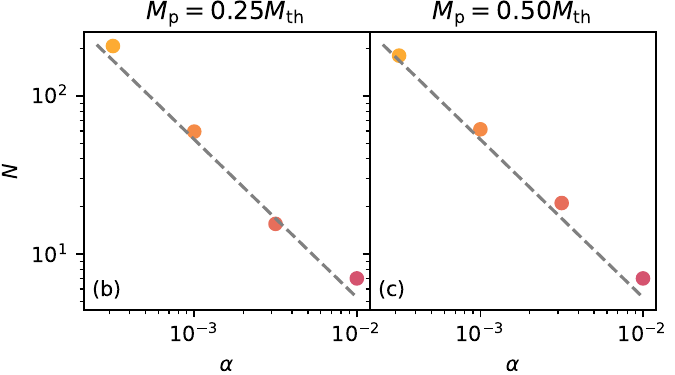}
    \caption{
    Top: Averaged normalised $\Sigma$ deviation measure $\Delta \sigma_\mathrm{av}$ defined by equation (\ref{eq:gap_deviation}) for viscous and inviscid simulations. Diamonds indicate the point of 15\% deviation from the inviscid (black) case. Grey dashed curve represents theoretical linear growth, see Section \ref{sec:initial_evolution}. Initially, the gap depths follow the inviscid linear prediction, but taper off with time, which occurs earlier for higher $\alpha$.
    Bottom: Number of planetary orbits $N$ that it takes for a viscous simulation with a given $\alpha$ to deviate by more than 15\% in $\Delta \sigma_\mathrm{av}$ from the inviscid curve, as a function of $\alpha$. Data are shown for $\Mp=0.25\Mth$ (b) and $\Mp=0.5\Mth$ (c).
    The gray line shows the timescale $t_{l-\nu}$ see equation (\ref{eq:viscous_time-scale}), which matches $N(\alpha)$ data well.}
    \label{fig:viscous_figure_2}
\end{figure}
%%%%%%%%%%%%%%%%%%%%%%%%%%%%%%%%%%%%%%%%%%%%%%%%%%

%%%%%%%%%%%%%%%%%%%%%%%%%%%%%%%%%%%%%%%%%%%%%%%%%%

\subsection{Comparison to viscous simulations}
\label{sec:visc_compare}

%%%%%%%%%%%%%%%%%%%%%%%%%%%%%%%%%%%%%%%%%%%%%%%%%%

We now verify the estimates obtained in Section \ref{sec:viscous_analytic} using a suite of viscous, globally-isothermal simulations, in which we add kinematic viscosity
in the form (\ref{eq:nu}) to the equations of motion. These runs used a surface density slope of $p = 1.5$ to ensure the background is initially in a viscous steady state, with  $\dot{M} = \text{const}$ and initial radial velocity $v_R = 3 \nu / (2 r)$. Viscous simulations are run for $\hp = 0.05, \Mp = {0.25, 0.5}$ and 
$\alpha \in \{ 10^{-2}, 10^{-2.5}, 10^{-3}, 10^{-3.5}, 10^{-4} \}$. The boundary conditions for $\Sigma$ and $\Omega$ are the same as in the inviscid case, but a radial velocity of $v_R = 3 \nu / (2 R)$ is now enforced at the boundaries. Further simulation details are available in Appendix \ref{sec:numerical_a}.

In Figure \ref{fig:split_up_sigma_viscous}, similarly to Figure \ref{fig:split_up_sigma}, we display the different terms of the surface density evolution equation (\ref{eq:time_evolve_sigma}) at two different moments of time based on a simulation with $\alpha = 10^{-3.5}$. One can see that early on, at $t=10\Pp$, the viscous term (green) is subdominant compared to the $\dot l$ term (orange), which itself is comparable to the $\fdep$ term (blue); this is similar to what one finds in the inviscid Figure \ref{fig:split_up_sigma}. However, things look very differently at $t=200\Pp$: the $\dot l$ term is subdominant compared with the viscous one, which is now comparable to the $\fdep$ contribution. Also, viscous term is noticeably non-zero even in the coorbital region $|R-\Rp|<\lsh$. This switchover can be naturally understood by noticing that $t_{l-\nu}\approx 170 \Pp$ for this run. Thus, panels (a) and (b) in Figure \ref{fig:split_up_sigma_viscous} correspond to still essentially inviscid ($t<t_{l-\nu}$) and viscously modified ($t>t_{l-\nu}$) regimes, correspondingly, see equation (\ref{eq:viscous_time-scale}). Since the amplitude of the $\dot l$ term decays with time compared to the $\fdep$ term, we conclude that viscosity slows down $\Sigma$ evolution in the process of gap opening and $\partial l/\partial t$ becomes small.

In Figure \ref{fig:viscous_figure_1} we show the time evolution of azimuthally averaged $\Sigma$ for different viscosities. One can see how the gap structure evolves from the inviscid case (black) progressively earlier as $\alpha$ increases, in agreement with equation (\ref{eq:viscous_time-scale}). For example, at 100 orbits the $\alpha = 10^{-4}$ case (lightest yellow) still follows the inviscid case very closely, whereas the $\alpha = 10^{-2}$ case has developed a flatter bottom in its shallow primary gap, which is no longer deepening at that point. Intermediate values of $\alpha$ fall cleanly in between these two curves. And by $200\Pp$ even $\alpha = 10^{-3}$ run shows a near flat-bottomed primary gap, while the $\alpha = 10^{-3.5}$ case still maintains a double gap structure. Both the $\alpha = 10^{-3.5}$ and $\alpha = 10^{-4}$ cases display both secondary and tertiary inner gaps, whereas the higher $\alpha$ cases only include either the primary and secondary gaps, with the exception of the $\alpha = 10^{-2}$ case where only the primary gap forms. This agrees with the results of \citet{miranda_gaps_2020}.

To further quantify the deviation of gap development in viscous discs from the inviscid case, we calculate in all runs the average gap deviation measure\footnote{We choose this particular metric to reduce the effect of the overall gap shape variation with $\alpha$.} $\Delta \sigma_\mathrm{av}$ defined by equation (\ref{eq:gap_deviation}) and follow its time evolution as a function of $\alpha$. We show $\Delta \sigma_\mathrm{av}(t)$ for the $\Mp = 0.25 \Mth$ case and different $\alpha$ in Figure \ref{fig:viscous_figure_2}a. Each of the $\Delta \sigma_\mathrm{av}(t)$ curves initially follow the inviscid curve (black) before tapering off. In the $\alpha = 10^{-2}$ case the $\Delta \sigma_\mathrm{av}$ measure is flattening out as the gap approaches a steady state in this high-viscosity disc.

We use this data to determine a time scale for the viscous effects to become dominant by measuring the number of orbits $N$ it takes for $\Delta \sigma_\mathrm{av}(t)$ to deviate from its inviscid counterpart by more than $15\%$ (somewhat arbitrarily chosen). These data are plotted in Figures \ref{fig:viscous_figure_2}b,c as a function of $\alpha$ for both $\Mp = 0.25 \Mth$ and $\Mp = 0.5 \Mth$. The number $N$ decays with $\alpha$ in very good agreement\footnote{This agreement is dependent on our choice of $15\%$ deviation; the slope of $N(\alpha)$ dependence is more important than its amplitude.} with out prediction (\ref{eq:viscous_time-scale}) for $t_{l-\nu}$ shown as the dashed line, and this behavior is independent of $\Mp$, as expected. 

%%%%%%%%%%%%%%%%%%%%%%%%%%%%%%%%%%%%%%%%%%%%%%%%%%
%%%%%%%%%%%%%%%%%%%%%%%%%%%%%%%%%%%%%%%%%%%%%%%%%%

\section{Discussion}
\label{sec:discussion}

%%%%%%%%%%%%%%%%%%%%%%%%%%%%%%%%%%%%%%%%%%%%%%%%%%

Our results demonstrate that understanding of gap opening in inviscid discs requires a very careful treatment of the angular momentum budget in the disc near the planet. More specifically, a key role is played by the time dependence of the specific angular momentum $l$ caused by its sensitivity to the pressure (or $\Sigma$) gradient (see equation (\ref{eq:Om_dev})), time-varying in the course of gap opening. This effect qualitatively changes the picture of gap opening, by allowing gas to be evacuated from the planetary coorbital region, where no deposition of the planet-driven density wave angular momentum takes place. Under the standard theory \citep{rafikov_planet_2002} this 
would not happen, see equation (\ref{eq:no-ldot-soln}).

Our results suggest the following physical picture of gap opening and mass evacuation from the vicinity of the planetary orbit when there is no angular momentum deposition within the shocking distance from the planet, i.e. $\fdep(R)=0$ for $|R-\Rp|<\lsh$. Let us first look at the vicinity of $R=\Rp+\lsh$ and neglect the background pressure gradient due to $\Sigma_0(R)$. Just outside of this radius $\fdep$ is nonzero, pushing gas away from the planet, which leads to $\partial P/\partial R<0$ around $\Rp+\lsh$. According to equation (\ref{eq:Om_dev}), this would {\it lower} the specific angular momentum $l$ around $R=\Rp+\lsh$, including just inside of this radius. But the total angular momentum of the fluid elements just inside of this $R$ has not changed, since $\fdep=0$ there. This means that these fluid elements must readjust their radial position to preserve their angular momentum and, since $\partial l/\partial R>0$, they must move {\it outwards}. This gives rise to an outward mass flux at $R=\Rp+\lsh$, resulting in mass flow out of the coorbital region. Similarly, gap opening just inside of $R=\Rp-\lsh$ leads to positive $\partial P/\partial R>0$ and locally {\it increases} $l(R)$, meaning that an {\it inward} gas flow must take place at this radius to preserve the total angular momentum of the gas; this again contributes to the evacuation of the coorbital ribbon of gas. As this evacuation starts first around $\Rp\pm\lsh$, it proceeds in an outside-in fashion and slower at $\Rp$, leading to a local peak of $\Sigma$ at $\Rp$ with $\Sigma(\Rp)$ steadily decreasing, see Figures \ref{fig:compare_same_mass} and \ref{fig:compare_diff_mass}.   

%%%%%%%%%%%%%%%%%%%%%%%%%%%%%%%%%%%%%%%%%%%%%%%%%%
\begin{figure}
    \centering
    \includegraphics[width=0.45\textwidth]{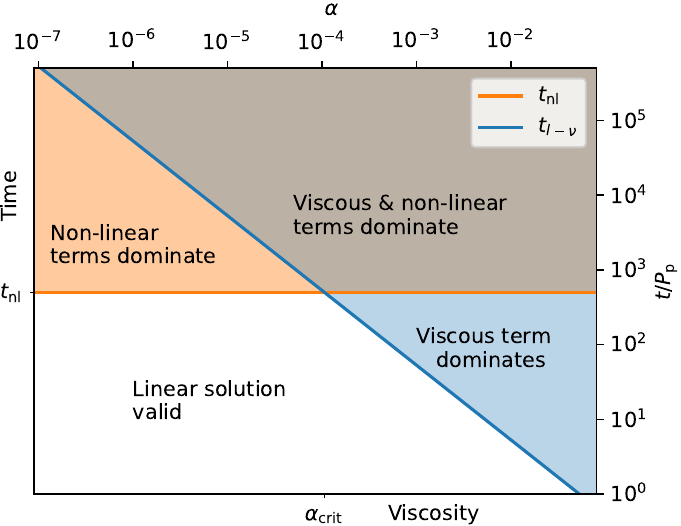}
    \caption{
    Schematic diagram of validity of the linear solutions obtained in Section \ref{sec:initial_evolution} as a function of time $t$ and dimensionless viscosity $\alpha$. The upper and lower axes illustrate a particular case of a disk with $\hp = 0.05$ and a planet mass of $\Mp/\Mth = 0.25$. The linear solution gets invalidated by the nonlinear effects for $t>t_\mathrm{nl}$ (see equations (\ref{eq:time_gap}), (\ref{eq:tnl_fixed})) and by viscosity for $t>t_{l-\nu}(\alpha)$ (see equation (\ref{eq:viscous_time-scale})). See text for details.}
    \label{fig:alpha_crit}
\end{figure}
%%%%%%%%%%%%%%%%%%%%%%%%%%%%%%%%%%%%%%%%%%%%%%%%%%

The importance of the $\dot l$ term in equation (\ref{eq:time_evolve_sigma}) has been previously recognized by \citet{belyaev_angular_2013}, \citet{Coleman2022} in their studies of the boundary layers of accretion discs and by \citet[][see their equations (20) and (21)]{arzamasskiy_disk_2018} who studied the effect of spiral shocks on accretion discs. 
This term was also mentioned qualitatively in the context of planet-disc interactions by  \cite{muto_two-dimensional_2010}. In their work they used simulation data to compute the components of the standard surface density equation, i.e. Equation (\ref{eq:base_mass}) without the $\dot l$ term (obtained by their explicit assumption that $\Omega$ does not deviative significantly from its unperturbed value),  and showed that the combination of these terms was unable to describe surface density evolution correctly.  Based on this, they concluded that $\Omega$ must evolve in time to drive mass flux in the co-orbital region.  This is in line with our results, which show that the $\dot l$ term plays a key role in describing surface density evolution even when the deviation from Keplerian rotation is small (e.g. at the first moment of gap opening,  see Section \ref{sec:orderofmaginitue}). Based on our results, we caution against neglecting this term without proper justification, especially in inviscid discs or in situations when disc properties vary rapidly.

Analytical solutions that we derived in Section \ref{sec:initial_evolution} and Appendix \ref{sec:app-non-loc} while accounting for the $\dot l$ cBased on our results, we caution against neglecting this term with-ontribution can find a variety of applications, see Section \ref{sec:applications} below. However, as they have been derived under the assumption of linearity in $\sigma=\delta\Sigma/\Sigma$, they apply only for a limited time, $t\lesssim t_\mathrm{nl}$, see the equations (\ref{eq:sig_max}) \& (\ref{eq:tnl_fixed}). Moreover, in discs with nonzero viscosity, our linear solutions may break down even earlier, as the results of Section \ref{sec:viscosity} demonstrate. Using equations (\ref{eq:time_gap}), (\ref{eq:tnl_fixed}), and (\ref{eq:viscous_time-scale}) one can determine a critical value of the dimensionless viscosity $\alpha_\mathrm{crit}$ for which $t_\mathrm{nl}=t_{l-\nu}$:
\begin{equation}
\label{eq:alpha_crit}
   \alpha_\mathrm{crit} \approx 1.5\,\hp\left(\frac{\Mp}{\Mth}\right)^{2} \left[B(\Mp/\Mth)\right]^{-1}.
\end{equation}
For $\alpha\gtrsim \alpha_\mathrm{crit}$ it is the viscosity rather than the nonlinear effects (see Appendix \ref{sec:validity}) that would invalidate our linear solutions first.

We show a schematic of this in Figure \ref{fig:alpha_crit}, which displays the range of applicability of our linear solutions from Section \ref{sec:initial_evolution} as a function of time $t$ and $\alpha$. The un-shaded (white) region is where the linear solutions do apply, while everywhere else some corrections are needed, either due to viscosity  (for $\alpha>\alpha_\mathrm{crit}$) or non-linearity (for $\alpha<\alpha_\mathrm{crit}$) becoming important. The upper/right axes of this figure illustrate a particular case of a planet-disc system with $\Mp/\Mth=0.25$, $\hp=0.05$, for which $t_\mathrm{nl}\approx 500\Pp$ and  $\alpha_\mathrm{crit} \approx 10^{-4}$. 

Next we discus applications (Section \ref{sec:applications}), limitations and extensions (Section \ref{sec:limitations}) of our work and compare it with the existing literature (Section \ref{sec:compare}).

%%%%%%%%%%%%%%%%%%%%%%%%%%%%%%%%%%%%%%%%%%%%%%%%%%

\subsection{Applications of our model}
\label{sec:applications}

%%%%%%%%%%%%%%%%%%%%%%%%%%%%%%%%%%%%%%%%%%%%%%%%%%

%%%%%%%%%%%%%%%%%%%%%%%%%%%%%%%%%%%%%%%%%%%%%%%%%%

\subsubsection{Observational implications}
\label{sec:observations}

%%%%%%%%%%%%%%%%%%%%%%%%%%%%%%%%%%%%%%%%%%%%%%%%%%
Our calculations can be directly applied to understanding the nature of planet-driven substructures in young, low viscosity discs, in which gap development may still be in its initial phases. For example, AS209 is a young ($\sim 1$ Myr old), low-mass ($M_\star=0.8 M_\odot$) star \citep{Andrews2018} with a protoplanetary disc that features a rich system of gaps and rings \citep{Huang2018}. Simulations by \citet{zhang_disk_2018} show that multiple annular substructures can be naturally explained as being due to a $\Mp/M_\star=10^{-4}$ ($\Mp\approx 27 M_\oplus$) at $\Rp=99$ AU, provided that the disc is thin, $\hp=0.05$, and almost inviscid, $\alpha=10^{-5}$. Adopting these parameters for illustration, we find $\Mp/\Mth=0.8$ and $\tgap\approx 0.5$ Myr, see equation (\ref{eq:time_gap}). Although this is shorter than the age of the system, the planet is likely younger than 1 Myr, and may easily be younger than $\tgap$, in which case its main gap would still be evolving. As $t_\mathrm{nl}\approx 0.036$ Myr, see equation (\ref{eq:tnl_fixed}), our simple linear solutions of Section \ref{sec:initial_evolution} would likely not be applicable (although this would change in hotter discs, e.g. for $h/r=0.07$ and the same $\Mp$ one finds $\tgap\approx 3.5$ Myr and $t_\mathrm{nl}\approx 0.26$ Myr, likely comparable to the planetary age). However, the $\dot l$ term should still be playing an important role in the gap evolution as the viscosity is very low, $t_{l-\nu}\approx 5.8$ Myr and $\alpha\ll \alpha_\mathrm{crit}\approx 2\times 10^{-3}$, see equations (\ref{eq:viscous_time-scale}) and (\ref{eq:alpha_crit}). 

Possible ongoing gap opening in AS209 may also be supported by the arrangement of substructures around 100 AU \citep{Huang2018}: two main peaks (B74 and B120) at 74 AU and 120 AU, a weak peak B97 at 97 AU, and two troughs (D90 and D105) at 90 AU and 120 AU, respectively. One may interpret B97 as being coorbital with the planet, two main peaks as residing at the pressure maxima, at $R_\mathrm{IR1}$, $R_\mathrm{OR1}$ where gas $\Sigma$ peaks, and two troughs as being at $R_\mathrm{IG1}$, $R_\mathrm{OG1}$ where $\Sigma$ is minimized, see Section \ref{sec:theorycomparison} and Figures \ref{fig:compare_same_mass},\ref{fig:compare_diff_mass}. We would then find $|R_\mathrm{IG1,OG1} -\Rp| \approx 7$ AU $\approx 1.5\,\lsh$, $|R_\mathrm{IR1,OR1}-\Rp| \approx 23$ AU $\approx 4.9\,\lsh,$ as $\lsh\approx 5$ AU for the adopted parameters. This is not too different from $1.6\,\lsh$ and $3.7\,\lsh$, respectively, predicted by our calculations according to the Figure \ref{fig:gas_spacings}a-d, suggesting that the system of substructures at 74-120 AU may indeed be compatible with the $\Sigma$ profile of an {\it evolving} gap, as determined in this work. 

Obviously, exactly matching the locations of all observed rings and gaps would require us to assess the possible role of nonlinear effects and to fully account for the details of dust dynamics, see Section \ref{sec:dust}. Disc thermodynamics are also quite important (see Section \ref{sec:th}), e.g. we use a globally isothermal EoS, while \citet{zhang_disk_2018} employ a locally isothermal EoS. Nevertheless, this exercise clearly shows that our semi-analytical analysis of evolving gaps can be used to directly interpret substructures observed in young, low-viscosity discs.

%%%%%%%%%%%%%%%%%%%%%%%%%%%%%%%%%%%%%%%%%%%%%%%%%%

\subsubsection{Application to dust dynamics}
\label{sec:dust}

%%%%%%%%%%%%%%%%%%%%%%%%%%%%%%%%%%%%%%%%%%%%%%%%%%

Disc substructures are most commonly found in sub-mm dust emission, and dust is known to move differently from gap. For example, it is known that rather modest ($10-20\%$, consistent with the constraint (\ref{eq:sig_max})), localized perturbations of gas $\Sigma$ --- a regime that we focus on in this work --- can lead to dramatic, order of magnitude variations of dust density $\Sigmad$ \citep{Dong2017}. This suggests that one could use the analytical results of Section \ref{sec:initial_evolution} and Appendix \ref{sec:full_global} for the time evolution of $\Sigma$ (or $\sigma$) in the linear regime to understand the appearance of observable dust substructures in protoplanetary discs without resorting to numerical simulations. This study would be relevant also for understanding the effect of pebble isolation \citep{lambrechts_rapid_2012, lambrechts_separating_2014, bitsch_pebble_2018} on planetary growth.  

Such a calculation can be accomplished by solving a 1D version of the continuity equation (\ref{eq:base_mass}) but for the dust surface density $\Sigmad$, which would use the radial velocity of the dust component $v_R=v_{R,\mathrm{d}}$ as an input. \citet{takeuchi_radial_2002} give
\begin{align}
v_{R,\mathrm{d}} &= v_R \frac{1}{\tau_s^2 + 1} - \eta \, v_\mathrm{K} \frac{\tau_s}{\tau_s^2 + 1},
\label{eq:vRd}
\end{align}
where $v_\mathrm{K}=\left(GM_\star/R\right)^{1/2}$ is the Keplerian velocity, $\eta = -(R \OmK^2 \Sigma)^{-1} (\partial P / \partial R)$ 
is the ratio of the gas pressure gradient to the stellar gravity, and $\tau_s$ is the dimensionless stopping time of the dust particles. We can use the equations (\ref{eq:vR}), (\ref{eq:amfd_func}), (\ref{eq:ldot}), (\ref{eq:lin-time}) to find that, e.g. in an inviscid disc the radial gas velocity is 
\begin{equation}
    v_R(R) \approx \frac{\fdep(R)}{\pi R^2 \OmK} - H^2 \,\frac{\partial \dot{\sigma}(R)}{\partial R},
    \label{eq:vr_sol}
\end{equation}
where we approximated $\partial l/\partial R\approx \OmK R/2$. Similarly, for the locally isothermal EoS the equations (\ref{eq:linearity}) \& (\ref{eq:lin-time}) give
\begin{equation}
    \eta(R,t) \approx h^2\left(s-R \,\frac{\partial \dot\sigma(R)}{\partial R}t\right),
    \label{eq:eta}
\end{equation}
where $s=-\partial \ln P_0/\partial \ln R$ is the background pressure gradient slope with $P_0=\Sigma_0 c_s^2$ ($s=p+2q$ for power law scalings (\ref{eq:Sig-PL}) \& (\ref{eq:c_s-PL})). 

The key simplifications allowed by the linear analytical solutions obtained in this work are (1) that the time dependence of $v_{R,\mathrm{d}}$ is simple, explicit and appears only through $\eta(R,t)$, and (2) that $\partial \dot\sigma/\partial R$ is a function of $R$ only that needs to be calculated only once using the solutions (\ref{eq:full_solution}) or (\ref{eq:global_start})-(\ref{eq:global_end}). A combination of the continuity equation and equations (\ref{eq:vRd})-(\ref{eq:eta}) allows one to efficiently study the dust structures forming in the vicinity of a low mass planet (as a function of dust size or $\tau_s$), something that will be explored in a future work.

%%%%%%%%%%%%%%%%%%%%%%%%%%%%%%%%%%%%%%%%%%%%%%%%%%

\subsection{Comparison to previous gap opening work}
\label{sec:compare}

%%%%%%%%%%%%%%%%%%%%%%%%%%%%%%%%%%%%%%%%%%%%%%%%%%

Development of planetary gaps has been previously followed in a number of numerical works, e.g. \citet{Yu2010}, \citet{zhu_low-mass_2013,Zhu2014}, \citet{Dong2017}, \citet{miranda_planet-disk_2020}. These studies invariably find the gap to have two minima, around $\lsh$ away on each side of the planetary orbit, with a maximum of $\Sigma$ --- a 'ribbon' --- in between, as predicted by \citet{rafikov_planet_2002}. At the same time, the ribbon gets steadily eroded over time (different from the blue curve in Figure \ref{fig:fig_one}b), contrary to the expectation of \citet{rafikov_planet_2002}, which is precisely the point addressed in our current work. Some of these simulations also reveal the self-similar, linear in $t$ character of the early stages of gap opening, see e.g \citet[][Fig. 9]{duffell_global_2012} or \citet[][Fig. 12c]{cimerman_planet-driven_2021}.

So far, to the best of our knowledge, only \cite{muto_two-dimensional_2010} and \cite{cimerman_emergence_2023} have looked into the initial stages of gap formation from a more (semi-)analytical perspective. Both studies connected gap formation to the generation of vortensity (specific vorticity) some distance away from the planet, which is caused by the planetary density waves evolving into shocks. These studies have showed that this process naturally results in the depletion of mass from the planetary coorbital region, something that we find here as well. Despite this similarity, their methods are \textbf{somewhat}  different from (although related to) our approach based on the angular momentum balance in the planetary vicinity. We provide a detailed comparison of our work with each of these studies in Appendix \ref{sec:gap_compafre}.

A number of past studies \citep[e.g.][]{duffell_gap_2013, duffell_empirically_2020, kanagawa_formation_2015, kanagawa_mass_2015} have looked into the structure of planet-driven gaps once they have reached a steady state, in which the injection of angular momentum by the planetary shock is exactly balanced by the viscous stress and $\Sigma$ is no longer evolving in time.  
We can compare our steady state solution (\ref{eq:vss_solution}) to these earlier studies.  In particular, in the limits of $R\ll\Rp$ and $R\gg\Rp$ our solution agrees with the equation (19) of \cite{dempsey_pileups_2020}. Additionally, we can compare our results with previous calculations of gap depth, by computing the right-hand side of the equation (\ref{eq:vss_solution}) for any $R$ satisfying $|R-\Rp|<\lsh$, e.g. for $R=\Rp$.  By noting that $\int_0^{R_p} \fdep(x) dx$ represents the one sided Linblad torque, we can write (using the equation (\ref{eq:Fdep}) and the boundary conditions $F_\mathrm{wave}(0)=F_\mathrm{wave}(\Rp)=0$)
\begin{equation}
  \Sigma_0 \int_0^{R_p} \fdep(x) dx = \int_0^{R_p} \frac{\partial \Fdep}{\partial R} dx = \int_0^{R_p} \frac{\partial T}{\partial R} dx \approx - 0.12 \pi F_{J, 0},
\end{equation}
where in the final equality we have chosen the same estimate for the one-sided Linblad torque as \cite{kanagawa_formation_2015, kanagawa_mass_2015} and $F_{J, 0}$ is given by Equation (\ref{eq:classic_torque}). Evaluating Equation (\ref{eq:vss_solution}) at the radial location of the planet with this anzatz gives
\begin{align}
    \sigma(R_p) &= - \, \frac{0.12 \pi}{3 \pi \alpha H^2  R_p^2 \Omega_p^2} \frac{F_{J, 0}}{\Sigma_p} 
    %&\approx - \, 0.04 \left(\frac{\Mp}{M_{\star}}\right)^{2} \left(\frac{\Hp}{\Rp}\right)^{-5} \alpha^{-1} 
    = - 0.04 K \, ,
    \label{eq:vss_gap_depth}
\end{align}
where $K = \left(\Mp/M_{\star}\right)^{2} \left(\Hp/\Rp\right)^{-5} \alpha^{-1}$ is the non-dimensional parameter defined in \cite{kanagawa_formation_2015}; one must have $K\ll 1$ to satisy the  assumption of shallow gaps.  In their analysis, and with numerical verification, \cite{kanagawa_formation_2015} found that the depth of the gap at $R_p$ is given by 
\begin{equation}
    \Sigma_{\mathrm{gap}}/\Sigma_0 = (1 + 0.04 K)^{-1}\approx 1-0.04K
\end{equation}
for $K\ll 1$, which matches our result (\ref{eq:vss_gap_depth}) exactly.

%%%%%%%%%%%%%%%%%%%%%%%%%%%%%%%%%%%%%%%%%%%%%%%%%%

\subsection{Limitations and extensions of our model}
\label{sec:limitations}

%%%%%%%%%%%%%%%%%%%%%%%%%%%%%%%%%%%%%%%%%%%%%%%%%%

Our linear model developed in Section \ref{sec:initial_evolution}, is constrained to work only for gaps of relative depth $ \lesssim 20\%$, however the importance of the $\dot l$ term in mediating planetary gap opening in low-$\alpha$ discs persists even for deeper gaps (see Figure \ref{fig:split_up_sigma}b), when nonlinear effects become important. We have focused primarily on the non-linear damping of planet-generated waves with $\fdep=0$ for $|R-\Rp|<\lsh$, and this is not a problem for viscous discs as non-zero $\alpha$ impacts wave damping only weakly \citep{miranda_gaps_2020}. However, in discs with linear radiative damping the radial profile of $\fdep$ will be significantly modified \citep {miranda_planet-disk_2020} and would be nonzero in the planetary coorbital region. However, this is not a problem since our general solutions (\ref{eq:lin-time}), (\ref{eq:full_solution}) and (\ref{eq:global_solution}) work for arbitrary $\fdep(R)$ (see also Section \ref{sec:time_dependent}). 

Similarly, even though we studied only the 2D discs and the form of $\fdep$ that we used (Appendix \ref{sec:deposition_models}) is based on 2D simulations, same approach can be used for fully 3D discs. One simply needs to come up with a prescription for $\fdep(R)$, based on theory or 3D simulations, that would properly describe the damping of the planet-driven density waves in 3D.

%%%%%%%%%%%%%%%%%%%%%%%%%%%%%%%%%%%%%%%%%%%%%%%%%%

\subsubsection{Role of disc thermodynamics}
\label{sec:th}

All semi-analytical calculations presented in this work used $\fdep$ determined for globally isothermal discs, see \citet{cimerman_planet-driven_2021} and Appendix \ref{sec:deposition_models}. At the same time, equation (\ref{eq:global_inviscid}) applies also to the locally isothermal EoS and Appendix \ref{sec:temp_gradient} provides a global solution (\ref{eq:global_start})-(\ref{eq:global_end}) for $\sigma$ in this case. However, to use this solution one still needs to provide a prescription for $\fdep(R)$ that is intrinsic to the locally isothermal discs. One curious feature of such discs is that they do not conserve the angular momentum flux of the freely-propagating waves even in the absence of non-linear wave damping. This implies that $\fdep$ would be non-zero even in the coorbital region, prior to wave shocking, driving non-zero mass flux there even in the absence of $\dot l$ term, see equation (71) in \citep{miranda_planet-disk_2020}. However, one can show (e.g. using our analytical solution (\ref{eq:blerg})) that this anomalous mass flux is formally subdominant (by a factor $\sim \hp$) compared to the one due to the $\dot l$ term.  

More sophisticated assumptions about disc thermodynamics, e.g. radiative $\beta$-cooling, would change not only $\fdep(R)$, but also the form of equation (\ref{eq:global_inviscid}) because a different treatment of the pressure gradient would be required in equation (\ref{eq:Om_dev}).

%%%%%%%%%%%%%%%%%%%%%%%%%%%%%%%%%%%%%%%%%%%%%%%%%%

\subsubsection{Extensions to time-dependent planet parameters}
\label{sec:time_dependent}

In our study we assumed that planetary characteristics --- $\Rp$ and $\Mp$ --- and disc properties stay fixed in time. In reality, planets can migrate, changing $\Rp$, and accrete the surrounding gas, increasing $\Mp$; on long timescales the disc would evolve, changing its $\Sigma$ and $c_s$. We can easily extend our model to include the planet's evolutionary effects by specifying a time-dependent angular momentum deposition model $\fdep(R) \rightarrow \fdep(R, t)$ in equation (\ref{eq:global_inviscid}). Noticing that $t$ would appear in this equation only as a parameter (in the source term with $\fdep$), it can be solved for the unknown $\dot{\sigma}$ exactly as before (see Section \ref{sec:initial_evolution} and Appendix \ref{sec:full_global}), but keeping in mind that now $\dot\sigma(R)\to \dot{\sigma}(R,t)$. For example, in the local limit (Section \ref{sec:local}) we would find 
\begin{align}
\dot{\sigma}(R, t) = 
  \, \frac{1}{2\pi}\frac{1}{\Hp^2 \Rp^2 \Omp} &\bigg[ e^{R/\Hp} \int_{0}^{R} \fdep(x,t) e^{ - x/\Hp} \md x 
\nonumber \\
& + e^{-R/\Hp} \int_{0}^{R} \fdep(x,t) e^{x/\Hp} \md x 
\nonumber \\
& -2 \sinh\left( R/\Hp \right) \int_{0}^{\infty} \fdep(x,t) e^{ - x/\Hp} \md x  \bigg], 
\label{eq:full_solution_time}
\end{align}
instead of equation (\ref{eq:full_solution}). The full solution is then obtained as
\begin{equation}
    \sigma(R, t) = \int_0^{t} \dot{\sigma}(R, t^\prime)\, \md t^\prime,
\end{equation}
instead of equation (\ref{eq:lin-time}). This general solution (with a properly specified angular momentum deposition model $\fdep(R,t)$) should again be accurate for gap depths $\lesssim 0.2 \Sigma_0$.

%%%%%%%%%%%%%%%%%%%%%%%%%%%%%%%%%%%%%%%%%%%%%%%%%%
%%%%%%%%%%%%%%%%%%%%%%%%%%%%%%%%%%%%%%%%%%%%%%%%%%

\section{Summary}
\label{sec:sum}

%%%%%%%%%%%%%%%%%%%%%%%%%%%%%%%%%%%%%%%%%%%%%%%%%%

In this work we have studied the early stages of gap development by planets in protoplanetary discs. Our key finding is that the time dependence of the specific angular momentum $l$ (or angular frequency $\Omega$) of the disc material in the planetary vicinity plays a crucial role in the angular momentum balance during gap opening. This time dependence is caused by the time-varying radial pressure support in the disc \citep{arzamasskiy_disk_2018}, which changes as the gap develops. This $\dot l$ contribution is responsible for steady mass evacuation from the planetary coorbital region --- a part of the disc where there is no angular momentum deposition by the the planetary density waves. We developed a linear approach to analytically study early gap opening with the $\dot l$ contribution fully incorporated, and obtained both local (Sections \ref{sec:local}-\ref{sec:time-scale}) and fully global (Appendix \ref{sec:app-non-loc}) solutions for gap evolution. 

These analytical solutions demonstrate that in inviscid discs gap opening starts as a self-similar process, with the radial profile of the gap having a universal shape and amplitude growing linearly in time, see equation (\ref{eq:lin-time}), which sets a characteristic gap opening time $\tgap$ (equation (\ref{eq:time_gap})). The radial profile of the surface density perturbation $\delta\Sigma$ is determined by the radial pattern of angular momentum deposition in the disc by the planetary density waves. 

We verified these predictions with inviscid numerical simulations over a range of disc and planetary parameters, finding excellent agreement for relative gap depths $\delta\Sigma/\Sigma\lesssim 20\%$; for deeper gaps nonlinear effects become important. We have also shown that even in viscous discs our linear solutions remain valid for a certain period of time $t_{l-\nu}$, see equation (\ref{eq:viscous_time-scale}). Our analytical approach is sufficiently flexible to potentially account for more sophisticated thermodynamic assumptions or for the evolution of planetary and disc parameters, e.g. planet migration and accretion.

These findings can be directly applied to interpret observations of substructures in young, low-viscosity protoplanetary discs. Our analytical results will also be used in the future to explore dust dynamics in the vicinity of young planets in a fast and efficient way. 

Finally, while we focused on the problem of gap opening by planets, we stress that the importance of the $\dot l$ term is a general result that may emerge in many other astrophysical disc settings.

\section*{Acknowledgements}

We thank Nicholas Cimmerman for useful discussions and support in this project and Jeremy Goodman for motivating R.R.R. to think about this problem. We also thank Takayuki Muto for his in-depth review of the paper, as well as for clarifying their previous results and for providing corrected mass fluxes from \cite{muto_two-dimensional_2010}.
A.J.C. is funded by the Royal Society of New Zealand Te Apārangi and the Cambridge Trust through the Cambridge-Rutherford Memorial Scholarship.
R.R.R. acknowledges financial support through the STFC
grant ST/T00049X/1 and the IAS.
Part of this work was performed using resources provided by the Cambridge Service for Data Driven Discovery (CSD3) operated by the University of Cambridge Research Computing Service (www.csd3.cam.ac.uk).
\textit{Software:} NumPy \citep{harris_array_2020}, SciPy \citep{virtanen_scipy_2020}, Matplotlib \citep{thomas_a_caswell_matplotlibmatplotlib_2023} and Athena++ \citep{stone_athena_2020}.

%%%%%%%%%%%%%%%%%%%%%%%%%%%%%%%%%%%%%%%%%%%%%%%%%%

\section*{Data Availability}
Simulation data will be made available upon reasonable request to the corresponding author.
Software implementations of the vortenstiy reconstruction algorithm, the angular momentum deposition function and the linear solutions for surface density evolution can be accessed on the authors GitHub under an open source license: \href{https://github.com/cordwella/vortensity\_evolution}{https://github.com/cordwella/vortensity\_evolution}.

An animated version of Figure \ref{fig:2d_visual} can be accessed as a part of the supplemental material.


% The best way to enter references is to use BibTe\sigma:

\bibliographystyle{mnras}
\bibliography{example} % if your bibtex file is called example.bib

%\begin{thebibliography}{99}
%\end{thebibliography}

%%%%%%%%%%%%%%%%%%%%%%%%%%%%%%%%%%%%%%%%%%%%%%%%%%

%%%%%%%%%%%%%%%%% APPENDICES %%%%%%%%%%%%%%%%%%%%%

\appendix

%%%%%%%%%%%%%%%%%%%%%%%%%%%%%%%%%%%%%%%%%%%%%%%%%%
%%%%%%%%%%%%%%%%%%%%%%%%%%%%%%%%%%%%%%%%%%%%%%%%%%

\section{Derivation of 1D fluid equations} 
\label{sec:3d}

%%%%%%%%%%%%%%%%%%%%%%%%%%%%%%%%%%%%%%%%%%%%%%%%%%

To ensure that our 1D equations (\ref{eq:base_mass})-(\ref{eq:base_amf}) are complete and accurate we re-derive them from the full 2D fluid equations. In polar $(R, \phi)$ coordinates,  
\begin{align}
    &\frac{\partial \Sigma}{\partial t} + \frac{1}{R} \frac{\partial}{\partial R} \left(R\Sigma v_R \right) + \frac{1}{R} \frac{\partial}{\partial \phi} \left( \Sigma v_{\phi} \right) = 0\,,
    \label{eq:cnt}
    \\
    &\Sigma \frac{\mathrm{D} \mathbf{v}}{\mathrm{D} t} = \Sigma \left[\frac{\partial \mathbf{v}}{\partial t} + (\mathbf{v} \cdot \nabla) \mathbf{v})\right] = - \nabla p + \nabla \cdot \boldsymbol{\tau} + \Sigma\, \nabla\Phi\,,
    \label{eq:NS}
\end{align}
where $\boldsymbol{\tau}$ is the viscous stress tensor and $\Phi$ is the total gravitational potential. We do this only for mass-weighed, azimuthally-averaged velocities defined by equation (\ref{eq:avv}), which we identify as variables in the 1D equations, as this leads to the simplest formulation of the problem. To derive conservation of angular momentum we consider the $\phi$ component of equation (\ref{eq:NS}),
\begin{equation}
    \Sigma \left(\frac{\partial v_{\phi}}{\partial t} + v_R \frac{\partial v_{\phi}}{\partial R} + \frac{v_{\phi}}{R} \frac{\partial v_{\phi}}{\partial \phi} + \frac{v_R v_{\phi}}{R}\right) = -\frac{1}{R}\frac{\partial p}{\partial \phi} + (\nabla \cdot \boldsymbol{\tau})_{\phi} - \frac{\Sigma}{R} \frac{\partial \Phi}{\partial \phi}.
\end{equation}
We can combine the equations (\ref{eq:cnt}), (\ref{eq:NS}) into the conservation of angular momentum equation:
\begin{align}
     R \frac{\partial}{\partial t}(\Sigma R v_{\phi}) 
     &+ \frac{\partial}{\partial R} \left(R^2 \Sigma v_R v_{\phi} \right)
     +  R \frac{\partial}{\partial \phi} \left( \Sigma v_{\phi}^2 \right)
     \nonumber\\
      &= - R \frac{\partial p}{\partial \phi} + R^2 (\nabla \cdot \boldsymbol{\tau})_{\phi} - R \Sigma \frac{\partial \Phi}{\partial \phi}.
      \label{eq:AMcont}
\end{align}

As viscous discs are not the primary consideration of this paper, we will provide only a simplified, approximate expression for the viscous contribution, assuming axisymmetric flow. In which case the only non-zero components of $\tau$ that enter equation (\ref{eq:AMcont}) are 
\begin{equation}
    \tau_{R \phi} = \mu \left( 
    \frac{\partial v_{\phi} }{\partial R } - \frac{ v_{\phi}}{R} \right)  = -\frac{G}{2 \pi R^2},
\end{equation}
where $\mu=\Sigma\nu$ and the viscous angular momentum flux $G$ has been defined after equation (\ref{eq:base_amf}). Then the divergence of the stress tensor becomes
\begin{align}
    (\nabla \cdot \boldsymbol{\tau})_{\phi} &= -\frac{1}{ 2 \pi R^2} \frac{\partial G}{\partial R}.
    \label{eq:div-tau}
\end{align}

We then apply the operator $\frac{1}{2\pi}\oint .. d\phi$ to the equations (\ref{eq:cnt}), (\ref{eq:AMcont}), using the mass-weighted definition (\ref{eq:avv}) of the averages:
\begin{align}
    &\frac{\partial \langle \Sigma \rangle }{\partial t} + \frac{1}{R} \frac{\partial}{\partial R} \left(R \langle \Sigma \rangle \langle v_R \rangle \right) = 0,
    \label{eq:cnt1}\\
     & R \frac{\partial}{\partial t}(R \langle \Sigma\rangle  \langle v_{\phi} \rangle ) 
     +  \frac{\partial}{\partial R} \left(R^2 \langle \Sigma \rangle  \langle v_R v_{\phi} \rangle  \right)
      = -\frac{1}{2 \pi} \left( \frac{\partial G}{\partial R} - \frac{\partial T}{\partial R} \right),
\label{tas}
\end{align}
where $\partial T/\partial R$ is the gravitational torque density defined by equation (\ref{eq:torque}), and we have used (\ref{eq:div-tau}).

With $\delta v_\phi \equiv v_\phi - \langle v_\phi \rangle$, 
the second term in equation (\ref{tas}) can be written as
\begin{align}
    R^2 \langle \Sigma \rangle  \langle v_R v_{\phi} \rangle  &= \frac{R^2}{2 \pi} \oint  \Sigma v_R (\langle v_{\phi} \rangle + \delta v_{\phi}) d\phi 
    \nonumber\\
    &= R^2 \langle v_{\phi} \rangle \langle\Sigma\rangle \langle v_R \rangle 
    + \frac{1}{2\pi} F_\mathrm{wave}, 
\end{align}
with $F_\mathrm{wave}$ is the wave angular momentum flux defined by equation (\ref{eq:f_wave}). Substituting this back into equation (\ref{tas}) we find
\begin{align}
     R \frac{\partial}{\partial t}(R \langle \Sigma\rangle  \langle v_{\phi} \rangle ) 
     &+  \frac{\partial}{\partial R} \left(R^2 \langle \Sigma \rangle  \langle v_R \rangle \langle v_{\phi} \rangle  \right)
     \nonumber\\
      &= -\frac{1}{2 \pi} \left( \frac{\partial G}{\partial R} - \frac{\partial T}{\partial R} + \frac{\partial F_\mathrm{wave}}{\partial R} \right).
      \label{tas1}
\end{align}
Using the definition (\ref{eq:Fdep}) of the deposition torque density and dropping angular brackets from all variables, equations (\ref{eq:cnt1}), (\ref{tas1}) finally reduce to the 1D equations (\ref{eq:base_mass})-(\ref{eq:base_amf}). 

Note that, with the exception of the viscous contribution, these equations are exact, thanks to our mass-weighted velocity definitions. This last aspect is critical, as the averages not weighted by $\Sigma$ do not result in a set of simple, closed-form equations. \cite{dempsey_pileups_2020} and \cite{muto_two-dimensional_2010} performed derivations with the unweighted (spatial) averages and had to make additional approximations to arrive at the 1D equations (\ref{eq:base_mass})-(\ref{eq:base_amf}). 

\cite{balbus_instability_1998} and \citet{Balbus1999} performed a  derivation similar to ours with mass-weighed averages to describe turbulent stresses in discs, but without planetary potential or wave transport. We note that our definition of $F_\mathrm{wave}$ would in general also include turbulent angular momentum flux if turbulence were present; in this work we assume the disc to be laminar.

%%%%%%%%%%%%%%%%%%%%%%%%%%%%%%%%%%%%%%%%%%%%%%%%%%
%%%%%%%%%%%%%%%%%%%%%%%%%%%%%%%%%%%%%%%%%%%%%%%%%%

\section{Angular Momentum Deposition Model} 
\label{sec:deposition_models}

%%%%%%%%%%%%%%%%%%%%%%%%%%%%%%%%%%%%%%%%%%%%%%%%%%

\begin{table}
\caption{Corrected fitting parameters for $\Delta \chi(\tau)$ in Equation (30) of \protect \citet{cimerman_planet-driven_2021}.}
\label{tab:fitting_parameters}
\begin{tabular}{llllll}
\hline
           & A    & $\tilde{\tau_b}$ & $\alpha_1$ & $\alpha_2$ & $\Delta$ \\
\hline
Inner Disc & $4.15$ & $0.402$            & $-7.80$      & $0.518$     & $0.595$    \\
Outer Disc & $6.19$ & $0.255$            & $-6.49$     & $0.540$      & $0.730$  \\
\hline
\end{tabular}
\end{table}

We use the density wave angular momentum deposition model from \cite{cimerman_planet-driven_2021} as an input to our analytical  calculations of the gap structure and evolution. That study verified the theory of \cite{rafikov_planet_2002} and derived semi-analytical prescriptions for the strength of shocks caused by planets as a function of the radial distance from the planet in globally isothermal discs, which were calibrated using direct 2D hydrodynamic simulations. \cite{cimerman_planet-driven_2021} considered shallow gaps, so that $\Sigma$ remains close to its unperturbed value given by equation (\ref{eq:Sig-PL}), and the amplitude of the density wave does not change as the gap grows, similar to our work. Here we briefly restate the main steps of their model in terms of angular momentum deposition (rather than vortensity).

Given $\Mp$, $\hp$, $\Rp$, the value of the angular momentum deposition function $\fdep(R)$ at a given $R$ is found via the following steps.

\begin{enumerate}

    \item Transform from $R$ to the space-like coordinate $\tau(R)$ using equation (18) of \citet{cimerman_planet-driven_2021}.

    \item Use this value of $\tau$ to compute $\tilde{\tau} = |\tau| - \tau_0 = |\tau| - 1.89 (\Mp/\Mth)$ and plug into equation (30) of \citet{cimerman_planet-driven_2021}, but with constant factors given in our Table \ref{tab:fitting_parameters}, to obtain the rescaled, dimensionless jump across the shock $\Delta \chi(\tau)$. 
    Here, we have re-derived anew the fitting parameters compared to Table 1 in \citet{cimerman_planet-driven_2021}, as an unfortunate scaling error made in that work caused their $\partial \Fdep/\partial R$ shown in their figures to be a factor of exactly $2 \pi$ larger than what would be calculated following their instructions.

    \item Convert from $\Delta \chi$ to the true density jump across the shock $\epsilon = (\Sigma - \Sigma_0)/\Sigma_0$
    using the modified version of equation (20) of \citet{cimerman_planet-driven_2021} as
    \begin{equation}
        \epsilon(R) = \Delta \chi \, \left(\frac{\Mp}{\Mth}\right)^{\theta}\left[\frac{|(R/\Rp)^{3/2} - 1|}{\sqrt{2} h_p (R/\Rp)^{-p+1}} \right]^{1/2}.
    \end{equation}
    While, theoretically, we would expect $\theta$ in this equation to be $1$ \citep[as in][]{cimerman_planet-driven_2021} when re-fitting the function $\Delta \chi$, we found that $\theta = 1.075$ was required to correctly match the mass dependence of the solutions shown in \citet{cimerman_planet-driven_2021}.
    \item Finally, we calculate the angular momentum deposition function using equation (19) of \citet{rafikov_protoplanetary_2016}:
    \begin{equation}
        \fdep(R) \equiv \frac{1}{\Sigma(R, t)}\frac{\partial \Fdep}{\partial R}  = \text{sign}[\Omp - \OmK(R)] \, m \, R\,  c_s^2 \,\psi_Q(\epsilon),
        \label{eq:fdep_prescr}
    \end{equation}
    with $m=1$ as appropriate for planetary density waves and $\psi_Q$ given as a function of $\epsilon$ by equation (28) in \citet{cimerman_planet-driven_2021} as appropriate in the isothermal limit.  
\end{enumerate}

%%%%%%%%%%%%%%%%%%%%%%%%%%%%%%%%%%%%%%%%%%%%%%%%%%
%%%%%%%%%%%%%%%%%%%%%%%%%%%%%%%%%%%%%%%%%%%%%%%%%%

\section{Surface density evolution in the non-local case}
\label{sec:app-non-loc}

%%%%%%%%%%%%%%%%%%%%%%%%%%%%%%%%%%%%%%%%%%%%%%%%%%

%%%%%%%%%%%%%%%%%%%%%%%%%%%%%%%%%%%%%%%%%%%%%%%%%%
\begin{figure}
        \centering
\includegraphics[width=0.48\textwidth]{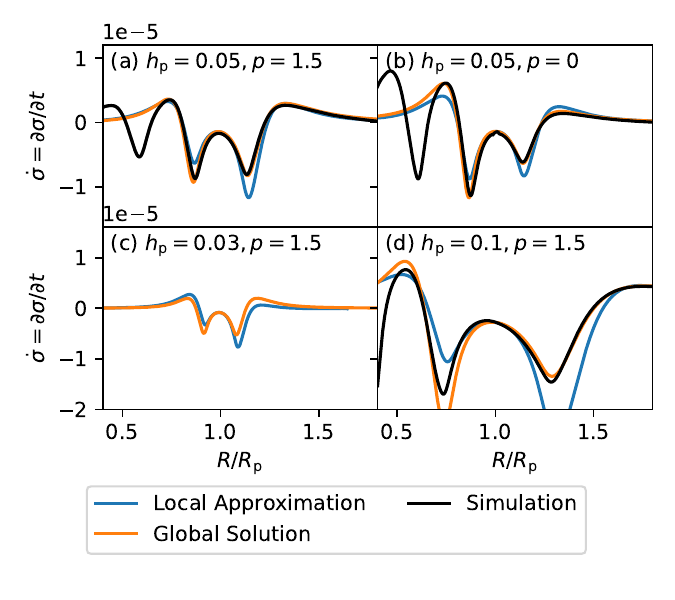}
    \caption{Comparison of local (equation \protect \ref{eq:local_dsigma}, blue) and global (equation \protect \ref{eq:full_solution}, orange) solutions for $\dot{\sigma}$, with globally isothermal simulations (black) for $\Mp = 0.25 \Mth$ at $t = 200 \Pp$ and 4 sets of disc parameters (labeled in each panel). The $\hp = 0.03$ case does not have a corresponding simulation --- we include it to show that the local and global solutions converge for lower $\hp$.
    There are systematic differences between the local and global solutions, e.g. the local one systematically overestimates (underestimates) $\dot\sigma$ in the outer (inner) gap. 
    }
    \label{fig:compare-local-global}
\end{figure}
\label{sec:full_global}
%%%%%%%%%%%%%%%%%%%%%%%%%%%%%%%%%%%%%%%%%%%%%%%%%%

Correctly matching our gap opening theory with simulations requires fully accounting for the background gradients in $\Sigma$, $\Omega$, etc. We assume a power law background surface density in the form (\ref{eq:Sig-PL}) and write out the $R$ dependence of $H$ in a globally isothermal disc as
\begin{equation}
    H = \frac{c_s}{\OmK} = \hp \Rp \left(\frac{R}{\Rp} \right)^{3/2}.
\end{equation}
We can then re-write the evolution equation (\ref{eq:global_inviscid}) as 
\begin{equation}
    \label{eq:global_full}
    \frac{\hp^2}{\Rp} \frac{\partial}{\partial R} \left[R^{4 - p} \frac{\partial \dot{\sigma}}{\partial R} \right] - R^{1 - p} \dot{\sigma} =  S(R),
\end{equation}
where $S(R)$ is the source term,
\begin{equation}
\label{eq:global_source}
    S(R) \equiv \frac{1}{\pi} \frac{\partial}{\partial R} \left[\frac{R^{-p-1}}{\OmK} \fdep(R) \right].
\end{equation}

This equation can be solved in terms of the modified Bessel functions $I_{\alpha}$ and $K_{\alpha}$ of order $\alpha$:
\begin{align}
    \sigma(R, t) & = 
    2\,t \, R^{(p-3)/2}\, \frac{\Rp}{\hp^2} \nonumber \\
    & \times \bigg[ K_{3 - p}\left(y(R)\right) \int_R^{\infty} I_{3 - p}\left(y(x)\right) x^{(p-3)/2} S(x) \md x  \nonumber \\
    & - I_{3 - p}\left( y(R) \right) \int_0^{R} K_{3 - p}\left(y(x) \right) x^{(p-3)/2} S(x) \md x 
    \bigg],
    \label{eq:global_solution}
\end{align}
where 
\begin{align}
y(x)=\frac{2}{\hp} \left(\frac{x}{\Rp}\right)^{-1/2}. 
\label{eq:y1}
\end{align}
It is this version of the solution that is used in the main body of the paper when comparing with numerical results. 

In Figure \ref{fig:compare-local-global} we compare the local solution for $\dot\sigma$ given by equation \ref{eq:local_dsigma}) with the global solution (\ref{eq:global_solution})-(\ref{eq:y1}) and simulations (gray), when available. All plots in that figure made for different $\hp$ or $p$ use the same $\fdep(R)$ described in Appendix \ref{sec:deposition_models}. Generally, the local solution predicts gap depths less accurately compared to the global solution which matches simulations better. As expected, the disagreement between the two solutions reduces as $\hp$ decreases. In addition, lowering $\hp$ improves the agreement of the global solution with the simulation (as expected, the two differ in the inner disc because of secondary arm forming). 

%%%%%%%%%%%%%%%%%%%%%%%%%%%%%%%%%%%%%%%%%%%%%%%%%%
\begin{figure}
    \centering
    \includegraphics[width=0.48\textwidth]{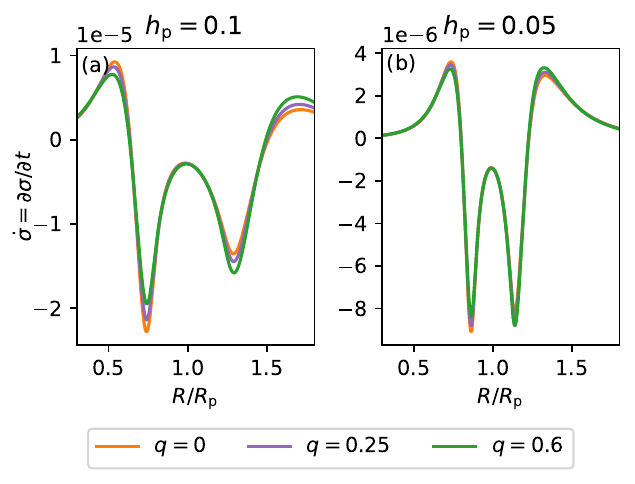}
    \caption{Comparison of $\dot \sigma$ profiles in discs with the same $\fdep(R)$ and locally isothermal thermodynamics. Different curves show the solution (\ref{eq:global_start})-(\ref{eq:global_end}) for 3 values of $q$, see equation (\ref{eq:c_s-PL}): 0, 0.25 and 0.6. The overall effect of varying $q$ is rather modest, especially for low $\hp$. See text for details.
    }
    \label{fig:compare-local-isothermality}
\end{figure}
%%%%%%%%%%%%%%%%%%%%%%%%%%%%%%%%%%%%%%%%%%%%%%%%%%

%%%%%%%%%%%%%%%%%%%%%%%%%%%%%%%%%%%%%%%%%%%%%%%%%%
%%%%%%%%%%%%%%%%%%%%%%%%%%%%%%%%%%%%%%%%%%%%%%%%%%

\subsection{Solution for a background temperature gradient}
\label{sec:temp_gradient}

%%%%%%%%%%%%%%%%%%%%%%%%%%%%%%%%%%%%%%%%%%%%%%%%%%

While in the main body of this work we considered exclusively globally isothermal discs due to the lack of angular momentum deposition models for discs with other thermodynamic profiles, our master equation (\ref{eq:global_inviscid}) is valid also for the locally isothermal EoS. Moreover, we are also able to solve it for $\sigma$ assuming a power law background temperature profile. Indeed, let us assume that the sound speed behaves as 
\begin{equation}
    c_s = c_{s, \mathrm{p}} \left( \frac{R}{R_p} \right)^{-q},
    \label{eq:c_s-PL}
\end{equation}
where $c_{s, \mathrm{p}}$ is the sound speed at $\Rp$ and $q$ is a (positive) constant. Then
\begin{equation}
    H = \frac{c_{s, \mathrm{p}} \Rp^{-q}}{\sqrt{G M_\star}} R^{3/2 - q} = A R^{3/2 - q},~~~~\mbox{with}~~~~A \equiv \hp \Rp^{-q-1/2}.
\end{equation}
Inserting this into the inviscid evolution equation (\ref{eq:global_inviscid}) we get 
\begin{align}
    A^2 \frac{\partial}{\partial R} \left[R^{4 - p -2q} \frac{\partial \dot{\sigma}}{\partial R} \right]- R^{1 -p}\dot{\sigma} =   S(R),
\end{align}
where $S(R)$ retains its previous definition (\ref{eq:global_source}). This equation differs only by the extra factor of $R^{-2q}$ in the first term compared to the globally isothermal equation (\ref{eq:global_solution}) and reduces to it when $q=0$. It can then be rearranged into 
\begin{align}
    A^2 R^{4 - p - 2q} \frac{\partial^2 \dot{\sigma}}{\partial R^2} + A^2 (4 - p - 2q) R^{3 - p - 2q} \frac{\partial \dot{\sigma}}{\partial R} - R^{1 - p}\dot{\sigma} = S(R).
\end{align}
For $q < 1/2$ this has the solution 
\begin{align}
    \dot{\sigma}(R) = \frac{R^{(p-3)/2 + q}}{A^2(q - 1/2)} \bigg[& I_{|\beta|} \left(y(R)\right) \int_{0}^{R} K_{|\beta|}\left(y(x)\right) x^{(p-3)/2 + q} S(x) \md x 
    \nonumber\\
    &- K_{|\beta|} \left( y(R) \right) \int_{R}^{\infty} I_{|\beta|}\left(y(x) \right) x^{(p-3)/2 + q} S(x) \md x
    \bigg],
    \label{eq:global_start}
\end{align} 
and for  $q > 1/2$
\begin{align}
    \dot{\sigma}(R) = \frac{R^{(p-3)/2 + q}}{A^2(q - 1/2)} \bigg[& I_{|\beta|} \left(y(R)\right) \int_{R}^{\infty} K_{|\beta|}\left(y(x)\right) x^{(p-3)/2 + q} S(x) \md x 
    \nonumber\\
    &- K_{|\beta|} \left( y(R) \right) \int_{0}^{R} I_{|\beta|}\left(y(x)\right) x^{(p-3)/2 + q} S(x) \md x,
    \bigg],
\end{align}
where now
\begin{align}
    \beta \equiv \frac{p + 2q -  3}{1 - 2 q}~~~~~~~\mbox{                  and } ~~~~~~~ y(x) \equiv \left| \frac{1}{q - 1/2}\frac{1}{A} x^{q - 1/2} \right|.
    \label{eq:global_end}
\end{align}
For $q = 1/2$ the solution involves exponentials in $R$.

Just to illustrate of the effect of varying $q$, in Figure \ref{fig:compare-local-isothermality} we plot this solution for different $q$ and $\hp$ using the globally isothermal angular momentum deposition function from Appendix \ref{sec:deposition_models}, with $\fdep(R)=0$ in the coorbital region $|R-\Rp|<\lsh$ (even though the true $\fdep$ would be different in locally isothermal discs). One can see that for a fixed $\fdep$ the dependence of $\sigma$ on $q$ is rather weak, but that it is stronger for larger $\hp$.

%%%%%%%%%%%%%%%%%%%%%%%%%%%%%%%%%%%%%%%%%%%%%%%%%%
%%%%%%%%%%%%%%%%%%%%%%%%%%%%%%%%%%%%%%%%%%%%%%%%%%

\section{Numerical Methods}
\label{sec:numerical_a}

%%%%%%%%%%%%%%%%%%%%%%%%%%%%%%%%%%%%%%%%%%%%%%%%%%

Athena++ solves the hydrodynamic equations, 
\begin{align}
    \frac{\partial \Sigma}{\partial t} + \nabla \cdot (\Sigma \mathbf{v}) &= 0, \\
    \frac{\partial (\Sigma \mathbf{v})}{\partial t} + \nabla \cdot (\Sigma \mathbf{v} \otimes \mathbf{v} + P \mathbf{I}) &= - \Sigma \nabla \Phi,
\end{align}
in conservative form using a Godunov scheme, where $P$ is the pressure, $\Sigma$ surface density and $\mathbf{v}$ is the velocity vector. Gravitational potential $\Phi$ is the sum of the gravitational potential of a central star, $\Phi_{\star} = - GM_\star/R$, and the gravitational potential of the planet $\Phi_\mathrm{p}$. To avoid singularities at the location of the planet, a fourth order smoothed potential \citep{dong_density_2011},
\begin{equation}
    \Phi_\mathrm{p} = - G\Mp \frac{d^2 + (3/2) r_\mathrm{s}^2}{\left(d^2 + r_\mathrm{s}^2\right)^{3/2}},
\end{equation}
is used, where $d = |\mathbf{R} - \mathbf{R}_\mathrm{p}|$ is the distance from the planet and $r_\mathrm{s} = 0.6 \Hp$ is the smoothing length. We neglect the indirect potential of the planet. To avoid spurious shocks, the planetary potential is ramped up over 10 $\Pp$ at the start of the simulation. 

We evaluate the hydrodynamics in a non-rotating frame centered on the star. The simulation domain covers the full $2 \pi$ in azimuth and $(0.2, 4.0)$ in $R$ with the planet placed at $\Rp = 1$ in code units. Damping zones are used for $R < 0.28$ and $R > 3.4$, and damp $v_R$ and $\Sigma$ to their initial values. The radial grid structure is logarithmic in $R$ and a resolution of $(N_{\phi}, N_{R}) = (7200, 3600)$ is used. Further details, including tests of this setup are available in \cite{cimerman_planet-driven_2021}.

%%%%%%%%%%%%%%%%%%%%%%%%%%%%%%%%%%%%%%%%%%%%%%%%%%

\section{Quantifying the validity of our assumptions and linear solutions}
\label{sec:validity}

%%%%%%%%%%%%%%%%%%%%%%%%%%%%%%%%%%%%%%%%%%%%%%%%%%

%%%%%%%%%%%%%%%%%%%%%%%%%%%%%%%%%%%%%%%%%%%%%%%%%%
\begin{figure}
    \centering
    \includegraphics[width=0.48\textwidth]{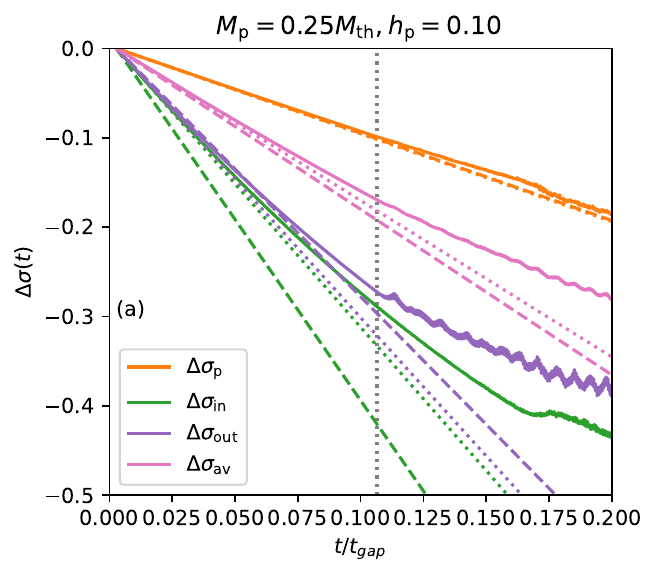}
    \includegraphics[width=0.48\textwidth]{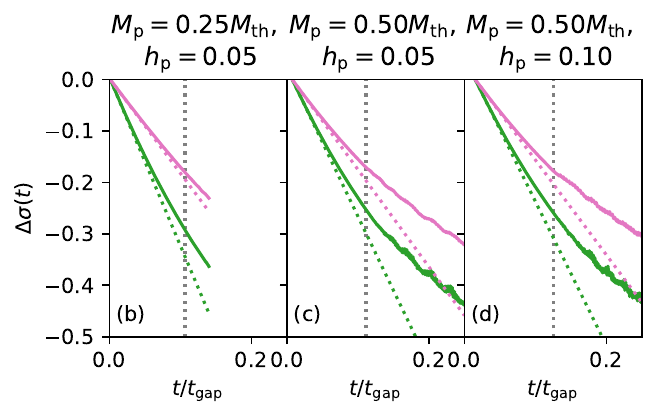}
    \caption{Different measures of gap depth as a function of time (in units of $\tgap$): $\sigma$ at the planetary location ($\Delta \sigma_\mathrm{p}$), at the nearest inner ($\Delta \sigma_\mathrm{in}$) and outer ($\Delta \sigma_\mathrm{out}$) gaps, as well as the average gap deviation ($\Delta \sigma_\mathrm{av}$) defined by equation (\ref{eq:gap_deviation}). We compare simulation (solid), linear theory, equation (\ref{eq:full_solution}) (dashed), and the
    linear (in $t$) evolution of $\Delta\sigma$ with  $\partial \Delta\sigma/\partial t$ measured in the simulation between 10 and 15 orbits. The gray vertical lines shows the time when the Rossby Wave instability sets in \protect \citep{cimerman_emergence_2023}. 
    Top: All $\Delta\sigma$ measures from a simulation with $\Mp/\Mth = 0.25, h_p=0.10, p=1.5$. 
    Bottom: $\Delta \sigma_\mathrm{in}$ and $\Delta \sigma_\mathrm{av}$ compared to simulation-calibrated linear evolution from the simulations with $(\Mp/\Mth, h_p, p) = \{
    (0.25, 0.05, 1.5), (0.50, 0.05, 1.5),
    (0.50, 0.10, 1.5) \}$.
     All simulations measures share the common trajectory of evolving initially linearly, before non-linear effects force them to slow down in evolution. See text for details.}
    \label{fig:gap_depths}
\end{figure}
%%%%%%%%%%%%%%%%%%%%%%%%%%%%%%%%%%%%%%%%%%%%%%%%%%

One approximation made in deriving the linear equation (\ref{eq:global_inviscid}) is that the relative perturbation of the surface density $\sigma\ll 1$, which in particular allowed us to obtain the result (\ref{eq:ldot}) and to neglect some higher order terms in equation (\ref{eq:global_inviscid}). As the depth of the gap linearly increases, at some point $\sigma$ will become large enough for this approximation to break down. Let us assume that this happens when the maximum of $\sigma$ reaches some characteristic value, $\max(\sigma) \approx \beta_\Sigma$, which we will determine later using simulations. We can then estimate the characteristic time $t_{\mathrm{nl}, \Sigma}$ for nonlinearity in $\Sigma$ to become important by adopting the solution (\ref{eq:blerg}), which should be at least approximately valid for $t<t_{\mathrm{nl}, \Sigma}$, and setting $\sigma=\beta_\Sigma$ and $R\approx \Rp\pm\lsh$ (as this is where the deepest part of the gap is). This exercise gives
\begin{equation}
\label{eq:tnl}
    t_{\mathrm{nl}, \Sigma} \approx \beta_\Sigma\, \tgap\, \left[ \cosh \left(\lsh/\Hp\right) \right]^{-1}
\end{equation} 
for the characteristic time when our linear solutions may start being inaccurate. 

Another assumption we made in Section \ref{sec:initial_evolution} is that $\Omega$ can be replaced with $\OmK$ (except in the $\dot l$ term). Let us denote $t_{\mathrm{nl}, \Omega}$ the characteristic time when this assumption gets violated upon $|\Omega^2 - \OmK^2|/\OmK^2$ reaching some characteristic tolerance $\beta_\Omega$. Using equation (\ref{eq:force_balance}), solution (\ref{eq:blerg}), and approximating $\Sigma^{-1} (\partial P/\partial R) \approx c_s^2(\partial \sigma/\partial R)$ we find
\begin{equation}
    \frac{\Omega^2 - \OmK^2}{\OmK^2} = \frac{t}{\tgap}\,\hp \,\sinh \left(\frac{R-\Rp}{\Hp}\right)
\end{equation}
in the coorbital region. Setting again $R\approx \Rp\pm\lsh$ we obtain
\begin{equation}
\label{eq:tnl_Om}
     t_{nl, \Omega} = \beta_\Omega\, \hp^{-1}\,\tgap \,\left[\sinh\left(\lsh/\Hp\right) \right]^{-1}.
\end{equation}
For small ($\Mp\lesssim \Mth$) planets $\lsh > \Hp$, and, assuming $\beta_\Sigma\sim\beta_\Omega$, one has $t_{\mathrm{nl},  \Omega} \sim \hp^{-1} t_{\mathrm{nl}, \Sigma}\gg t_{\mathrm{nl}, \Sigma}$ (as $\sinh y \approx \cosh y$ for $y \gtrsim 1$). This means that our linear solutions always become inaccurate because of $\delta\Sigma/\Sigma$ becoming nonlinear at $t\approx t_{\mathrm{nl}, \Sigma}$, and not because of $\Omega$ deviating from $\OmK$ by too much.

%%%%%%%%%%%%%%%%%%%%%%%%%%%%%%%%%%%%%%%%%%%%%%%%%%
\begin{table*}
\caption{List of parameters and various derived characteristics for the $p = 1.5$ inviscid simulations. Time to RWI onset, $t_\mathrm{RWI}$, is taken from \protect \cite{cimerman_emergence_2023}. We also display the analytical gap opening timescale $\tgap$ (equation (\ref{eq:time_gap})), the time $t_\mathrm{nl}$ for the inner gap depth $\Delta\sigma_\mathrm{in}$ to deviate from linear growth by $10 \%$, and deviation metrics for $\Sigma$ and $\Omega$.}
\label{tab:time_scale}
\begin{tabular}{rrrrrrrrr}
\hline
\multicolumn{1}{l}{$\Mp/\Mth$} & \multicolumn{1}{l}{$\hp$} & \multicolumn{1}{l}{$\tgap$} & \multicolumn{1}{l}{Time to reach 10\% deviation}           & \multicolumn{1}{l}{$t_\mathrm{nl}/\tgap$} & \multicolumn{1}{l}{Maximum inner gap}              & \multicolumn{1}{l}{Maximum Value of}                    & \multicolumn{1}{l}{Time to RWI } & \multicolumn{1}{l}{$t_\mathrm{RWI}/\tgap$} \\
\multicolumn{1}{l}{}             & \multicolumn{1}{l}{}      & \multicolumn{1}{l}{}                 & \multicolumn{1}{l}{from linearity, $t_\mathrm{nl}$} & \multicolumn{1}{l}{}                       & \multicolumn{1}{l}{depth at $t_\mathrm{nl}$, $\Delta\sigma_\mathrm{av}$} & \multicolumn{1}{l}{$|\Omega-\OmK|/\OmK$ at $t_\mathrm{nl}$} & \multicolumn{1}{l}{onset, $t_\mathrm{RWI}$}          & \multicolumn{1}{l}{}                  \\
\multicolumn{1}{l}{}             & \multicolumn{1}{l}{}      & \multicolumn{1}{l}{$(\Pp)$}          & \multicolumn{1}{l}{$(\Pp)$}                          & \multicolumn{1}{l}{}                       & \multicolumn{1}{l}{}             & \multicolumn{1}{l}{}                                    & \multicolumn{1}{l}{$(P_p)$}   & \multicolumn{1}{l}{}                  \\
\hline
0.1                              & 0.05                      & 113313                               & \textgreater{}7000                                   & -                                          & -                                            & -                                                       & \textgreater{}7000            & -                                     \\
0.1                              & 0.07                      & 80938                                & \textgreater{}5000                                   & -                                          & -                                            & -                                                       & \textgreater{}5000            & -                                     \\
0.1                              & 0.1                       & 56656                                & \textgreater{}3000                                   & -                                          & -                                            & -                                                       & \textgreater{}3000            & -                                     \\
0.25                             & 0.05                      & 7550                                 & 522                                                  & 0.073                                      & 0.210                                        & 0.108                                                   & 760                           & 0.104                                 \\
0.25                             & 0.07                      & 5390                                 & 376                                                  & 0.072                                      & 0.210                                        & 0.152                                                   & 540                           & 0.103                                 \\
0.25                             & 0.1                       & 3780                                 & 256                                                  & 0.072                                      & 0.210                                        & 0.220                                                   & 380                           & 0.104                                 \\
0.5                              & 0.05                      & 1540                                 & 98                                                   & 0.075                                      & 0.192                                        & 0.078                                                   & 146                           & 0.112                                 \\
0.5                              & 0.07                      & 1010                                  & 69                                                  & 0.075                                      & 0.185                                        & 0.109                                                   & 112                           & 0.120                                 \\
0.5                              & 0.1                       & 779                                  & 51                                                   & 0.078                                      & 0.190                                        & 0.156                                                   & 82                            & 0.126                   \\             
\hline
\end{tabular}
\end{table*}
%%%%%%%%%%%%%%%%%%%%%%%%%%%%%%%%%%%%%%%%%%%%%%%%%%

We now test these expectations with simulations. As equation (\ref{eq:tnl}) predicts $t_{\mathrm{nl}, \Sigma}$ to be some fraction of $\tgap$, we will use simulations to measure the 'nonlinearity timescale' $t_\mathrm{nl}$ at which our linear solutions for $\sigma$ deviate by a certain margin from what is found in simulations. We can use different metrics in this exercise, for example, the gap depth at the location of the planet ($\Delta \sigma_\mathrm{p}$), or at the deepest part of the closest inner ($\Delta \sigma_\mathrm{in}$) and outer ($\Delta \sigma_\mathrm{out}$) gaps. To account for the more global characteristics of $\sigma(R,t)$ profile, we also introduce an 'integrated' measure $\Delta \sigma_\mathrm{av}$ defined as
\begin{equation}
    \Delta \sigma_\mathrm{av} = \frac{1}{4 \lsh} \int_{\Rp -  2\lsh}^{\Rp + 2 \lsh} \sigma \md R,
    \label{eq:gap_deviation}
\end{equation}
where the integration range is chosen such that $\sigma<0$, predominantly, see Figures \ref{fig:fig_one}b,\ref{fig:compare_same_mass},\ref{fig:compare_diff_mass}; $\Delta \sigma_\mathrm{av}$ is less sensitive to slight changes in the gap structure than the other metrics.  

We plot these different measures in Figure \ref{fig:gap_depths}a, by showing simulations as solid curves and analytical predictions with the dashed curves of corresponding color. One can see that initially the two types of curves follow each other while decaying linearly with $t$, as expected from our linear solution, with the exception of $\Delta \sigma_\mathrm{in}$, for which the analytical and numerical curves have somewhat different slope from the start. This difference is caused by a systematic overestimate of the depth of the inner gap by our global solution (\ref{eq:global_solution}), see e.g. Figure \ref{fig:compare_same_mass}f. To avoid this issue while focusing on understanding the departure of gap evolution from the linear (in $t$) track (rather than worrying about the accuracy of the radial profile of analytical $\sigma(R,t)$), we also perform a second semi-numerical estimate of different gap depths. In this case we measure $\partial \sigma/\partial t$ from simulations between $10\Pp$ and $15\Pp$ and use it to linearly extrapolate the gap depths as a function of time. These are shown as dotted lines in Figure \ref{fig:gap_depths}, and provide a considerable improvement in the case of $\Delta \sigma_\mathrm{in}$.

Comparing solid and dotted curves, one can see that all gap depth measures from simulations are initially approximately linear and agree with the dotted lines, but begin to slowly level off at $t/\tgap$ increases and eventually exhibit oscillations due to the onset of the Rossby Wave Instability \citep{cimerman_emergence_2023}. The inner and outer gap measures deviate from linearity the earliest, whereas $\Delta \sigma_\mathrm{p}$ is still near linear even after the activation of the RWI.

In the end, we have adopted $\Delta \sigma_\mathrm{in}$ as our metric of choice, and defined $t_\mathrm{nl}$ as the time at which the inner gap depth deviates in our simulations by more than $\beta=10 \%$ from the initial linear trend. Table \ref{tab:time_scale} shows this  $t_\mathrm{nl}$ along with other relevant timescales and deviation measures for a range of simulated discs/planets. One can see that for all our simulations the $10\%$ deviation from linear theory occurred when $\Delta \sigma_\mathrm{in} \approx 0.2$. When time-scales are placed into units of $\tgap$, the time $t_\mathrm{nl}$ at which this relative deviation is reached does not depend on $\hp$ or $\Mp/\Mth$ and is about $0.07\tgap$. Also, $|\Omega-\OmK|/\OmK$ at $t=t_\mathrm{nl}$ is quite small, the departure of $\Omega$ from $\OmK$ $\lesssim 10-20\%$. These observations are summarized in the end of Section \ref{sec:theorycomparison}, see equations (\ref{eq:sig_max}) \& (\ref{eq:tnl_fixed}).

%%%%%%%%%%%%%%%%%%%%%%%%%%%%%%%%%%%%%%%%%%%%%%%%%%
%%%%%%%%%%%%%%%%%%%%%%%%%%%%%%%%%%%%%%%%%%%%%%%%%%

\section{Literature comparison}
\label{sec:gap_compafre}

%%%%%%%%%%%%%%%%%%%%%%%%%%%%%%%%%%%%%%%%%%%%%%%%%%

%%%%%%%%%%%%%%%%%%%%%%%%%%%%%%%%%%%%%%%%%%%%%%%%%%
\begin{figure}
    \centering
    \includegraphics[width=0.48\textwidth]{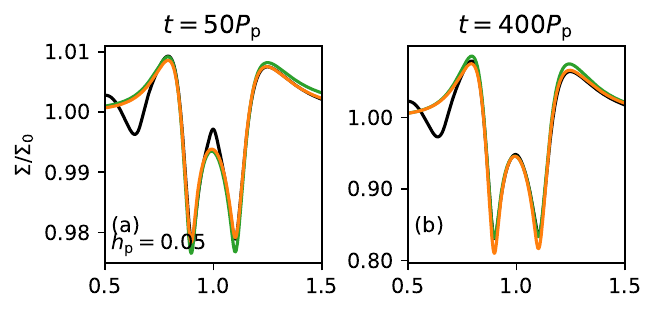}
    \includegraphics[width=0.48\textwidth]{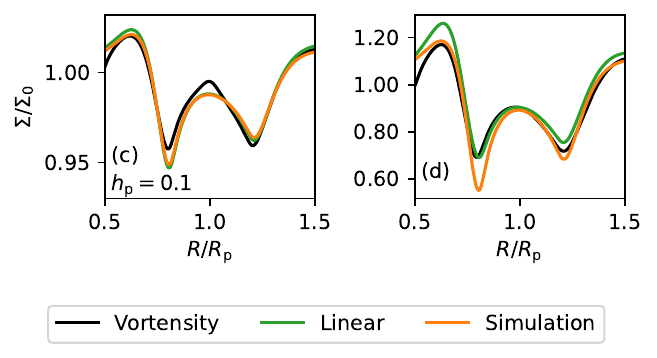}
    \caption{Comparison of gap properties calculated using our linear solutions (orange), the vortensity reconstruction method of \protect \cite{cimerman_emergence_2023} (blue), and fully non-linear numerical simulations (green) at two different moments of time $t=50\Pp$ (left) and $400\Pp$ (right). Top row: $\Mp/\Mth = 0.25, h_p = 0.05, p=1.5$; bottom row: $\Mp/\Mth = 0.25, h_p = 0.10, p=1.5$. The linear and vortensity-based methods agree well at early times, however later on the linear model deviates as it does not capture some nonlinear effects that the vortensity-based model accounts for.
    }
    \label{fig:vortensity_compare}
\end{figure}
%%%%%%%%%%%%%%%%%%%%%%%%%%%%%%%%%%%%%%%%%%%%%%%%%%

%%%%%%%%%%%%%%%%%%%%%%%%%%%%%%%%%%%%%%%%%%%%%%%%%%

\subsection{Comparison to \texorpdfstring{\citet{cimerman_emergence_2023}}{Cimerman \&Rafikov 2023}}
\label{sec:CR}

%%%%%%%%%%%%%%%%%%%%%%%%%%%%%%%%%%%%%%%%%%%%%%%%%%

\citet{cimerman_emergence_2023} developed a semi-analytical method for computing gap profiles following \citet{Lin2010}. This method considers the injection of vortensity (potential vorticity) 
$\zeta =(R \Sigma)^{-1} \md(R^2 \Omega)/\md R$ in an axisymmetric disc by the planetary shock, leading to an overall linear growth of vortensity as
\begin{equation}
    \zeta(R, t) = \zeta(R, t=0) + t S_{\zeta, \mathrm{sh}}(R),
    \label{eq:zeta_growth}
\end{equation}
where $S_{\zeta, \mathrm{sh}}$ is the vortensity source term for which a semi-analytical prescription was developed in \citet{cimerman_planet-driven_2021}. 
At any moment of time, once $\zeta(R, t)$ is specified via (\ref{eq:zeta_growth}), the surface density profile is obtained by numerically solving the non-linear ordinary differential equation
\begin{equation}
    \label{eq:vor_construction}
    \frac{1}{R^3}\frac{\md}{\md R}\left(R^3 \frac{\md \ln\Sigma}{\md R}\right) + \frac{\OmK}{c_s^2} = \zeta\frac{2\Sigma}{c_s^2}\left(\OmK^2 + \frac{c_s^2}{R}\frac{\md \ln\Sigma}{\md R}\right)^{1/2}.
\end{equation}
If we set $\Sigma = \Sigma_0(1 + \sigma)$ and linearize equation (\ref{eq:vor_construction}) in terms of $\sigma\ll 1$ assuming the local approximation, it can be reduced to a form analogous to equation (\ref{eq:local_dsigma}). This demonstrates a close correspondence between the vortensity reconstruction technique of \cite{cimerman_emergence_2023} and our current framework. However, at later times the method of \cite{cimerman_emergence_2023} may be more accurate because equation (\ref{eq:vor_construction}) incorporates some nonlinear effects.

\cite{cimerman_emergence_2023} and our work both use the same model for the planetary wave damping. Therefore by comparing our studies with the same $\fdep$ we can directly compare the different methods of constructing gap profiles. We compare $\Sigma(R)$ using our linear solution (\ref{eq:global_solution}), their calculation based on equation (\ref{eq:vor_construction}), and simulations in Figure \ref{fig:vortensity_compare}, for two different $\hp$ and moments of time. The two semi-analytical methods agree with each very well at early times. At later time the agreement noticeably worsens in the higher $\hp$ case (panel (d)) but is still good in the low $\hp$ case (panel (b)). This is because $t_\mathrm{nl}\approx 250\Pp$ for $\hp=0.1$, while $t_\mathrm{nl}\approx 500\Pp$ for $\hp=0.05$, meaning that in panel (d) our linear solution is no longer accurate because of nonlinear effects. Quite generally, the vortensity reconstruction technique provide a somewhat better match to $\Sigma(R)$ from simulations (aside from the secondary gap in the inner disc and small over-density at $\Rp$ at early times in simulations, which is an initial transient).

On the other hand, the linear solutions developed in this work are more practical at early times, while the condition (\ref{eq:tnl_fixed}) holds, than the method of \cite{cimerman_emergence_2023} in a few key ways. Firstly, our model is linear in time and does not require the numerical solution of a non-linear differential equation at every $t$, it can therefore be evaluated and manipulated with much greater ease by computing $\dot\sigma(R)$ just once and then using equation (\ref{eq:lin-time}). Secondly, the vortensity reconstruction method of \cite{cimerman_emergence_2023} is not easily amenable to the inclusion of viscosity.

%%%%%%%%%%%%%%%%%%%%%%%%%%%%%%%%%%%%%%%%%%%%%%%%%%

\subsection{Comparison to \texorpdfstring{\citet{muto_two-dimensional_2010}}{Muto et. al (2010)}}
\label{sec:Muto}

\cite{muto_two-dimensional_2010} studied planetarty gap opening by considering the second order perturbations to the background hydrodynamical equations in the shearing sheet approximation. They have shown that a localized production of vortensity (e.g. by the planet-driven shock) can drive a mass flux globally, including the planetary coorbital region where vortensity production is absent. They obtained an explicit expression for the mass flux driven by the vortensity production only for a model in which the vortensity source is composed of two Dirac $\delta$-functions,
\begin{equation}
    \label{eq:muto_source}
    S(r) = S_0 [ \delta(r - x_\mathrm{s}) - \delta(r + x_\mathrm{s}) ],
\end{equation}
where $S_0$ and $x_\mathrm{s}$ are positive constants and $r = R - \Rp$; vortensity production occurs only at $R=\Rp\pm x_\mathrm{s}$, and in our case we can associate $x_\mathrm{s}\to \lsh$. 
\cite{muto_two-dimensional_2010} provided an explicit expression for the mass flux in their equation (54) which lead to infinite mass fluxes at  $R=\Rp\pm x_\mathrm{s}$. This, however, was caused by algebraic errors in the evaluation of their equations (49)-(54) (Muto, private communication). 
A corrected evaluation of these equations provides a mass flux of 
\begin{equation}
    \mathcal{F}_{\mathrm{M}} = \frac{\Hp S_0}{c_s^2} e^{-l_{sh}/\Hp} \sinh(r/\Hp),
\end{equation}
(Muto, private communication) and therefore a surface density evolution:
\begin{equation}
\label{eq:muto_sol}
    \frac{\partial \Sigma}{\partial t} = - \, \frac{S_0}{c_s^2} e^{-l_{sh}/\Hp} \cosh(r/\Hp),
\end{equation}
which has the same form as our solution for the symmetric angular momentum deposition (Equation \ref{eq:most_simplifed_x}) with  $\psi(x) = \delta(x - 1)$.

\bsp	% typesetting comment
\label{lastpage}
\end{document}